\documentclass[12pt,a4paper]{article}

\usepackage[T1]{fontenc}
\usepackage[margin=1in]{geometry}
\usepackage{amsmath,amssymb,amsfonts}
\usepackage{bm}
\usepackage{graphicx}
\usepackage{booktabs}
\usepackage{array}
\usepackage{tabularx}
\usepackage{multirow}
\usepackage{float}
\usepackage{caption}
\usepackage{subcaption}
\usepackage{siunitx}
\usepackage{cite}
\usepackage{xcolor}
\usepackage{hyperref}
\usepackage{enumitem}
\usepackage{authblk}
\usepackage{microtype}
\usepackage{algorithm}
\usepackage{algpseudocode}
\usepackage{cleveref}

\setlist[itemize]{
	leftmargin=1.6em,
	itemsep=2pt,
	topsep=3pt
}

\newcolumntype{Y}{>{\centering\arraybackslash}X}

\crefname{algorithm}{algorithm}{algorithms}
\Crefname{algorithm}{Algorithm}{Algorithms}

\newcommand{\safeincludegraphics}[3][]{%
	\IfFileExists{#2}{%
		\includegraphics[#1]{#2}%
	}{%
		\fbox{%
			\parbox[c][5.0cm][c]{0.92\linewidth}{%
				\centering
				#3\\[0.5em]
				\footnotesize
				Figure generated by the accompanying numerical code.%
			}%
		}%
	}%
}

\title{
	Hard-Constrained Physics-Informed Neural Network with Adaptive Regional
	Residual Balancing for the Generalized Falkner--Skan Problem
}

\author[]{Mehari Fentahun Endalew}
\author[]{Xiaoming John Zhang}

\affil[]{
	Beijing Institute of Mathematical Sciences and Applications (BIMSA),
	Beijing, China
}

\date{}

\begin{document}
	
	\maketitle
	
	
	\begin{abstract}
		
		We present a physics-informed neural solver for the generalized Falkner–Skan boundary-value problem that combines an exact boundary-admissible trial representation with adaptive regional residual balancing (ARRB). The three prescribed boundary conditions are embedded analytically, eliminating boundary-condition penalty terms while allowing the finite-domain streamfunction value and wall shear to be determined by the governing equation. The residual is divided into wall, middle, and tail regions, and a physical global residual is reconstructed from regional mean-squared errors using region-length fractions. Safeguarded exponential-moving-average inverse-gradient coefficients adaptively balance the regional contributions during Adam optimization. A deterministic 800-point residual monitor is used for checkpoint selection, followed by two deterministic L-BFGS stages minimizing the physical global residual.
		The method is tested on the Blasius, favorable-pressure-gradient Falkner–Skan, and Pohlhausen cases. For \((\beta_0,\beta_1)=(0.75,0.50)\), the predicted wall shear is \(f''(0)=0.8997161394\), compared with the finite-domain reference \(0.8997168085\), giving an absolute error of \(6.691\times10^{-7}\). The residual MSE is \(6.357\times10^{-9}\), while the relative \(L_2\) errors in \(f'\) and \(f''\) are \(2.848\times10^{-6}\) and \(4.188\times10^{-5}\). A matched single-seed ablation shows that ARRB reduces residual MSE by 52.81\% relative to global-residual training and by 50.24\% relative to equal-regional weighting. Wall-shear errors are reduced by 77.30\% and 72.37\%, respectively. These results support ARRB as an accuracy-oriented residual-conditioning strategy, while multi-seed experiments remain necessary to quantify optimization variability.

	\end{abstract}
	
	
	
	\section{Introduction}
	\label{sec:introduction}
	
	Similarity solutions of the laminar boundary-layer equations provide a
	classical framework for investigating the interaction of viscous diffusion,
	convection, and pressure gradients. The Blasius equation describes the
	zero-pressure-gradient flat-plate boundary layer \cite{blasius1908},
	whereas the Falkner--Skan equation extends the similarity formulation to
	wedge flows and related external-velocity distributions
	\cite{falkner1931,schlichting2017}.
	
	Although the Falkner--Skan problem is one-dimensional, its nonlinear
	third-order structure, semi-infinite physical domain, and sensitivity of
	the wall quantity \(f''(0)\) make it a useful benchmark for nonlinear
	numerical methods \cite{asaithambi2021,temimi2018,abbasbandy2021}. The
	wall shear is particularly sensitive to the near-wall solution, whereas
	the normalized velocity approaches its outer value asymptotically. These
	features make the problem suitable for examining both boundary-condition
	enforcement and spatially nonuniform residual optimization.
	
	Physics-informed neural networks (PINNs) approximate solutions of
	differential equations by minimizing residuals constructed through
	automatic differentiation \cite{raissi2019}. Neural-network-based
	approaches have also been explored for Blasius and Falkner--Skan-type
	equations, including direct machine-learning implementations,
	orthogonal-polynomial neural blocks, and comparisons with recurrent and
	fully connected neural architectures
	\cite{narain2021,aghaei2023,anitha2024}. PINN optimization may, however,
	become difficult when different loss components or different spatial
	portions of the governing-equation residual generate substantially
	different parameter-gradient magnitudes
	\cite{wang2021,mcclenny2023}. Boundary penalties can introduce an
	additional source of competition because governing-equation accuracy must
	then be balanced against approximate boundary-condition satisfaction.
	
	Boundary-satisfying neural trial functions provide an established
	alternative in which prescribed boundary conditions are incorporated
	directly into the admissible solution space \cite{lagaris1998,mcfall2009}.
	Related hard-constraint formulations include constrained expressions,
	functional-connection methods, distance-function constructions, and
	hard--soft combinations designed to strengthen or exactly impose selected
	physical constraints
	\cite{leake2020,lu2021,sukumar2022,nguyen2026}.
	
	Hard boundary constraints and adaptive residual weighting address
	different optimization issues. Exact boundary admissibility removes the
	need to balance governing-equation and boundary-condition penalty terms.
	It does not, however, ensure that different spatial portions of the
	remaining governing-equation residual contribute comparably to parameter
	updates. Adaptive PINN strategies therefore seek to mitigate imbalance by
	modifying the relative influence of losses or residual information using,
	for example, gradient or residual information
	\cite{wang2021,mcclenny2023}. The present formulation applies this
	general motivation to spatially aggregated components of a single
	governing-equation residual.
	
	Recent work has also extended ansatz-based hard constraints to
	multidomain interface problems. Chung et al.~\cite{chung2026} considered
	hard-constrained PINN formulations for elliptic interface equations. The
	generalized Falkner--Skan problem considered here is structurally
	different: it is a single-domain, third-order boundary-value problem with
	no internal material interface or prescribed jump condition. The factor
	\(G(s)\) introduced below therefore acts as a global
	boundary-admissibility mask for a single neural correction rather than an
	interface-localized construction.
	
	Accordingly, the contribution claimed here is not the general concept of
	hard-constraint embedding itself. Instead, the proposed method combines a
	Falkner--Skan-specific admissible trial representation with adaptive
	regional residual balancing. The trial representation satisfies
	\[
	f(0)=0,\qquad f'(0)=0,\qquad f'(L)=1
	\]
	identically while retaining the degrees of freedom required for the
	governing equation to determine the wall shear \(f''(0)\) and the
	unprescribed terminal value \(f(L)\).
	
	The motivation for regional balancing follows from the spatially
	nonuniform Falkner--Skan solution structure. The near-wall region is
	particularly important for recovery of \(f''(0)\), whereas the outer
	region approaches \(f'(\eta)\rightarrow1\). A single scalar
	domain-averaged residual does not explicitly reveal whether different
	spatial regions generate substantially different optimization gradients,
	even when the boundary conditions themselves are already satisfied
	exactly.
	
	The computational domain is therefore partitioned into wall, middle, and
	tail regions. Regional residual mean-squared errors are used in two
	distinct ways. First, a physical domain-averaged residual is reconstructed
	using the corresponding region-length fractions. Its physical weighting
	is therefore independent of the integer number of stochastic collocation
	points assigned to each region. Second, safeguarded EMA-smoothed inverse
	proxy-gradient coefficients are used during Adam optimization to balance
	the regional residual components. These adaptive coefficients are
	optimization quantities and are distinct from the physical region-length
	weights.
	
	The principal contributions of this work are:
	\begin{enumerate}
		\item a Falkner--Skan-specific hard-boundary trial representation that
		enforces \(f(0)=0\), \(f'(0)=0\), and \(f'(L)=1\) identically while
		preserving the freedom required to determine \(f''(0)\) and \(f(L)\)
		from the governing equation;
		
		\item an adaptive regional residual balancing strategy in which wall,
		middle, and tail components of the governing-equation residual are
		balanced using safeguarded, EMA-smoothed inverse proxy-gradient
		coefficients;
		
		\item a physical domain-averaged residual constructed explicitly from
		regional residual means using region-length fractions, so that its
		physical weighting is independent of regional collocation density; and
		
		\item a hybrid computational protocol combining fresh stratified Adam
		collocation batches, deterministic fixed-grid checkpoint selection,
		physical-global-residual L-BFGS refinement with stagewise rollback,
		and quantitative validation against finite-domain BVP references and
		an analytical Pohlhausen wall-shear benchmark.
	\end{enumerate}
	
	The method is evaluated for the Blasius case, a selected
	favorable-pressure-gradient Falkner--Skan case, and the Pohlhausen
	benchmark. A matched single-seed residual-objective ablation is additionally
	used to isolate the effect of adaptive regional weighting while retaining
	the same hard-constrained trial representation and optimization protocol.
	
	The remainder of the paper is organized as follows. Section~2 introduces
	the generalized Falkner--Skan problem and benchmark cases. Section~3
	develops the hard-constrained trial representation, regional residual
	decomposition, adaptive weighting rule, and optimization objectives.
	Section~4 summarizes the computational algorithm. Section~5 describes the reference solutions and validation metrics. Section~6 gives them computational configuration. Section~7 presents the numerical results and
	residual-objective ablation. Section~8 discusses the numerical behavior
	and limitations of the method, and Section~9 gives the conclusions.
	
	
	\section{Generalized Falkner--Skan problem}
	\label{sec:problem}
	
	The generalized Falkner--Skan equation considered in this study is
	
	\begin{equation}
		f'''(\eta)
		+\beta_0 f(\eta)f''(\eta)
		+\beta_1\left[1-f'(\eta)^2\right]
		=0,
		\qquad
		\eta\in[0,L].
		\label{eq:general_fs}
	\end{equation}
	
	The finite-domain boundary conditions are
	
	\begin{equation}
		f(0)=0,
		\qquad
		f'(0)=0,
		\qquad
		f'(L)=1.
		\label{eq:boundary_conditions}
	\end{equation}
	
	A truncated domain \(L=12\) is used throughout the reported calculations
	to approximate the far-field condition. Here, \(\eta\) is the similarity
	coordinate, \(f(\eta)\) is the dimensionless streamfunction variable,
	\(f'(\eta)\) is the normalized velocity, and \(f''(0)\) is the
	dimensionless wall-shear parameter under the adopted similarity scaling.
	
	For the conventional one-parameter Falkner--Skan family,
	
	\begin{equation}
		\beta_0=\frac{m+1}{2},
		\qquad
		\beta_1=m,
		\label{eq:m_parameters}
	\end{equation}
	
	where \(m\) is the pressure-gradient parameter. With
	
	\begin{equation}
		c=\sqrt{\frac{m+1}{2}},
		\qquad
		\xi=c\eta,
		\qquad
		F(\xi)=cf(\eta),
	\end{equation}
	
	\cref{eq:general_fs} transforms to
	
	\begin{equation}
		F'''+FF''
		+\beta_H\left[1-(F')^2\right]=0,
		\qquad
		\beta_H=\frac{2m}{m+1}.
		\label{eq:hartree}
	\end{equation}
	
	The normalized velocity is unchanged because \(F'(\xi)=f'(\eta)\).
	
	The Blasius case is
	
	\begin{equation}
		(\beta_0,\beta_1)=(0.5,0),
		\label{eq:blasius_case}
	\end{equation}
	
	the selected favorable-pressure-gradient Falkner--Skan case is
	
	\begin{equation}
		(\beta_0,\beta_1)=(0.75,0.50),
		\label{eq:selected_case}
	\end{equation}
	
	which corresponds to \(m=0.5\), and the Pohlhausen benchmark is
	
	\begin{equation}
		(\beta_0,\beta_1)=(0,1).
		\label{eq:pohlhausen_case}
	\end{equation}
	
	The Pohlhausen pair does not belong to the one-parameter family in
	\cref{eq:m_parameters}. It nevertheless provides an  independent benchmark
	because an analytical semi-infinite-domain solution is available for this parameter pair \cite{pohlhausen1921}.
	
	
	\section{Hard-constrained PINN with adaptive regional residual balancing}
	\label{sec:method}
	
	
	\subsection{Neural representation and hard boundary constraints}
	\label{subsec:hard_representation}
	
	The hard-constrained formulation is based on the general principle of
	constructing a neural trial space in which the prescribed boundary
	conditions are satisfied identically. This idea has a long history in
	neural differential-equation solvers, where the trial solution is
	decomposed into a non-trainable component satisfying the prescribed
	boundary data and a trainable correction that preserves those conditions
	\cite{lagaris1998,mcfall2009}.
	
	For an ordinary differential equation with linear boundary operators
	
	\begin{equation}
		\mathcal{B}_j[f]=g_j,
		\qquad
		j=1,\ldots,m,
		\label{eq:general_bc_operator}
	\end{equation}
	
	a boundary-admissible representation may be written as
	
	\begin{equation}
		\hat f_\theta(\eta)
		=
		b(\eta)
		+
		\psi_\theta(\eta),
		\label{eq:general_hard_representation}
	\end{equation}
	
	where the lifting function \(b\) satisfies
	
	\begin{equation}
		\mathcal{B}_j[b]=g_j,
		\qquad
		j=1,\ldots,m,
	\end{equation}
	
	and the trainable correction satisfies
	
	\begin{equation}
		\mathcal{B}_j[\psi_\theta]=0,
		\qquad
		j=1,\ldots,m.
		\label{eq:homogeneous_correction}
	\end{equation}
	
	By linearity,
	
	\begin{equation}
		\mathcal{B}_j[\hat f_\theta]
		=
		\mathcal{B}_j[b]
		+
		\mathcal{B}_j[\psi_\theta]
		=
		g_j.
	\end{equation}
	
	Thus, the boundary conditions are incorporated into the trial space rather
	than enforced approximately through boundary-loss penalties.
	
	For the three-condition boundary set
	
	\begin{equation}
		f(0)=A,
		\qquad
		f'(0)=B,
		\qquad
		f'(L)=C,
		\label{eq:general_three_bcs}
	\end{equation}
	
	one admissible lifting is
	
	\begin{equation}
		b_{A,B,C}(\eta)
		=
		A+B\eta
		+
		(C-B)
		\frac{
			\eta-\left(1-e^{-\eta}\right)
		}{
			1-e^{-L}
		}.
		\label{eq:general_lifting_function}
	\end{equation}
	
	Its derivative is
	
	\begin{equation}
		b_{A,B,C}'(\eta)
		=
		B
		+
		(C-B)
		\frac{1-e^{-\eta}}{1-e^{-L}}.
	\end{equation}
	
	Hence,
	
	\begin{equation}
		b_{A,B,C}(0)=A,
		\qquad
		b_{A,B,C}'(0)=B,
		\qquad
		b_{A,B,C}'(L)=C.
	\end{equation}
	
	For the Falkner--Skan problem,
	
	\begin{equation}
		A=0,
		\qquad
		B=0,
		\qquad
		C=1,
	\end{equation}
	
	so that the lifting becomes
	
	\begin{equation}
		\boxed{
			b(\eta)
			=
			\frac{
				\eta-\left(1-e^{-\eta}\right)
			}{
				1-e^{-L}
		}}
		.
		\label{eq:base_function}
	\end{equation}
	
	A fully connected feed-forward neural network \(N_\theta(x)\) is used,
	with architecture
	
	\begin{equation}
		[1,100,100,1].
		\label{eq:network_architecture}
	\end{equation}
	
	Hyperbolic-tangent activations are used, and all neural-network
	calculations and automatic differentiation are performed in 64-bit
	floating-point precision.
	
	The physical coordinate is normalized according to
	
	\begin{equation}
		s=\frac{\eta}{L},
		\qquad
		x=2s-1,
		\qquad
		x\in[-1,1].
		\label{eq:coordinate_mapping}
	\end{equation}
	
	Define
	
	\begin{equation}
		H(s)=3s^2-2s^3
		\label{eq:endpoint_mode}
	\end{equation}
	
	and
	
	\begin{equation}
		G(s)=Ls^2(1-s)^2.
		\label{eq:neural_mask}
	\end{equation}
	
	The Hermite endpoint mode satisfies
	
	\begin{equation}
		H(0)=0,
		\qquad
		H(1)=1,
		\qquad
		H_s(0)=H_s(1)=0,
		\label{eq:hermite_endpoint_properties}
	\end{equation}
	
	whereas the neural mask satisfies
	
	\begin{equation}
		G(0)=G(1)=0,
		\qquad
		G_s(0)=G_s(1)=0.
		\label{eq:mask_endpoint_properties}
	\end{equation}
	
	The hard-constrained Falkner--Skan approximation is
	
	\begin{equation}
		\boxed{
			\hat f_\theta(\eta)
			=
			b(\eta)
			+
			c_\theta H(s)
			+
			G(s)N_\theta(x)
		}
		\label{eq:trial_solution}
	\end{equation}
	
	where \(c_\theta\) is a trainable scalar parameter and \(N_\theta\) is the raw neural-network output. In the implementation, \(c_\theta\) is stored as a model parameter together with the neural-network parameters.
	
	For the lifting function,
	
	\begin{equation}
		b(0)=0,
		\qquad
		b'(0)=0,
		\qquad
		b'(L)=1,
		\label{eq:base_endpoint_properties}
	\end{equation}
	
	because
	
	\begin{equation}
		b'(\eta)
		=
		\frac{1-e^{-\eta}}{1-e^{-L}}.
		\label{eq:base_derivative}
	\end{equation}
	
	Since
	
	\begin{equation}
		\frac{dH}{d\eta}
		=
		\frac{1}{L}H_s
	\end{equation}
	
	and
	
	\begin{equation}
		\frac{d}{d\eta}
		\left[
		G(s)N_\theta(x)
		\right]
		=
		\frac{1}{L}G_s(s)N_\theta(x)
		+
		\frac{2}{L}G(s)N_{\theta,x}(x),
	\end{equation}
	
	the endpoint properties imply
	
	\begin{equation}
		\boxed{
			\hat f_\theta(0)=0,
			\qquad
			\hat f_\theta'(0)=0,
			\qquad
			\hat f_\theta'(L)=1
		}
		\label{eq:hard_bcs}
	\end{equation}
	
	identically for all trainable parameters, apart from floating-point
	roundoff.
	
	The construction retains the second-derivative freedom required by the
	third-order equation. In particular,
	
	\begin{equation}
		G_{ss}(0)
		=
		G_{ss}(1)
		=
		2L.
		\label{eq:mask_second_derivative}
	\end{equation}
	
	Because \(G=G_s=0\) at the endpoints,
	
	\begin{equation}
		\left.
		\frac{d^2}{d\eta^2}
		\left[
		G(s)N_\theta(x)
		\right]
		\right|_{\eta=0}
		=
		\frac{2}{L}N_\theta(-1),
		\label{eq:mask_wall_second}
	\end{equation}
	
	and
	
	\begin{equation}
		\left.
		\frac{d^2}{d\eta^2}
		\left[
		G(s)N_\theta(x)
		\right]
		\right|_{\eta=L}
		=
		\frac{2}{L}N_\theta(1).
		\label{eq:mask_tail_second}
	\end{equation}
	
	Similarly,
	
	\begin{equation}
		H_{ss}(0)=6,
		\qquad
		H_{ss}(1)=-6.
		\label{eq:hermite_second_derivative}
	\end{equation}
	
	Moreover,
	
	\begin{equation}
		b''(\eta)
		=
		\frac{e^{-\eta}}{1-e^{-L}},
	\end{equation}
	
	so
	
	\begin{equation}
		b''(0)
		=
		\frac{1}{1-e^{-L}}.
	\end{equation}
	
	The wall curvature is therefore
	
	\begin{equation}
		\boxed{
			\hat f_\theta''(0)
			=
			\frac{1}{1-e^{-L}}
			+
			\frac{6c_\theta}{L^2}
			+
			\frac{2}{L}N_\theta(-1)
		}
		\label{eq:wall_shear_decomposition}
	\end{equation}
	
	and remains a trainable, equation-determined quantity.
	
	Since \(H(1)=1\) and \(G(1)=0\),
	
	\begin{equation}
		\hat f_\theta(L)
		=
		b(L)+c_\theta,
	\end{equation}
	
	while
	
	\begin{equation}
		\hat f_\theta'(L)=1.
	\end{equation}
	
	Thus, the unprescribed terminal streamfunction value \(f(L)\) also remains
	free.
	
	All linear layers are first initialized using Xavier-normal weights and
	zero biases. The weights and bias of the final neural-network layer are
	then reset to zero while the hidden-layer initialization is retained.
	The endpoint coefficient is initialized as \(c_\theta=0\). Consequently,
	the raw network output initially satisfies \(N_\theta(x)=0\) for every
	input \(x\), and the complete trial solution begins exactly from the
	analytical lifting profile,
	
	\begin{equation}
		\hat f_\theta(\eta)\big|_{\mathrm{initial}}
		=
		b(\eta).
	\end{equation}
	
	\subsection{Physics-informed residual}
	\label{subsec:residual}
	
	Automatic differentiation provides
	
	\begin{equation}
		\hat f_\theta',
		\qquad
		\hat f_\theta'',
		\qquad
		\hat f_\theta'''.
	\end{equation}
	
	The pointwise Falkner--Skan residual is
	
	\begin{equation}
		R_\theta(\eta)
		=
		\hat f_\theta'''
		+
		\beta_0
		\hat f_\theta
		\hat f_\theta''
		+
		\beta_1
		\left[
		1-
		\left(
		\hat f_\theta'
		\right)^2
		\right].
		\label{eq:residual}
	\end{equation}
	
	No boundary-condition penalty terms are included because
	\cref{eq:hard_bcs} holds analytically.
	
	
	\subsection{Regional residual decomposition}
	\label{subsec:regions}
	
	The computational domain is divided into
	
	\begin{equation}
		\Omega_{\mathrm{w}}
		=
		[0,3],
		\qquad
		\Omega_{\mathrm{m}}
		=
		(3,8),
		\qquad
		\Omega_{\mathrm{t}}
		=
		[8,12].
		\label{eq:regions}
	\end{equation}
	In the implementation, a point at \(\eta=3\) is assigned to the wall region and a point at \(\eta=8\) is assigned to the tail region, consistent with the interval convention in \cref{eq:regions}. For a collocation subset \(\mathcal{C}_r\) in region \(r\),
	
	\begin{equation}
		\mathcal{L}_r
		=
		\frac{1}{|\mathcal{C}_r|}
		\sum_{\eta_i\in\mathcal{C}_r}
		R_\theta(\eta_i)^2,
		\qquad
		r\in
		\{
		\mathrm{w},
		\mathrm{m},
		\mathrm{t}
		\}.
		\label{eq:regional_loss}
	\end{equation}
	
	The physical region fractions are
	
	\begin{equation}
		\omega_{\mathrm{w}}
		=
		\frac{3}{12}
		=
		\frac14,
	\end{equation}
	
	\begin{equation}
		\omega_{\mathrm{m}}
		=
		\frac{8-3}{12}
		=
		\frac{5}{12},
	\end{equation}
	
	and
	
	\begin{equation}
		\omega_{\mathrm{t}}
		=
		\frac{12-8}{12}
		=
		\frac13.
	\end{equation}
	
	The physical domain-averaged residual is therefore
	
	\begin{equation}
		\boxed{
			\mathcal{L}_{\mathrm{g}}
			=
			\frac14
			\mathcal{L}_{\mathrm{w}}
			+
			\frac{5}{12}
			\mathcal{L}_{\mathrm{m}}
			+
			\frac13
			\mathcal{L}_{\mathrm{t}}
		}.
		\label{eq:global_loss}
	\end{equation}
	
	Equation~\cref{eq:global_loss} is implemented directly. The residual MSE
	is first computed independently within each of the three regions, after
	which the regional means are combined using the corresponding physical
	region-length fractions. The physical global residual is therefore not
	computed as the unweighted mean over the complete stratified stochastic
	batch. Consequently, changing the integer number of collocation points
	assigned to a region does not change that region's prescribed physical weight in \(\mathcal{L}_{\mathrm{g}}\).
	
	\subsection{Stratified collocation sampling}
	\label{subsec:sampling}
	
	Each reported Adam step uses
	
	\begin{equation}
		N_c=500
	\end{equation}
	
	collocation points.
	
	Before integer rounding,
	
	\begin{equation}
		500
		\left(
		\frac14,\frac5{12},\frac13
		\right)
		=
		(125,\ 208.33\ldots,\ 166.67\ldots).
	\end{equation}
	
	A largest-remainder allocation gives
	
	\begin{equation}
		N_{\mathrm{w}}=125,
		\qquad
		N_{\mathrm{m}}=208,
		\qquad
		N_{\mathrm{t}}=167.
		\label{eq:regional_counts}
	\end{equation}
	
	The number of Adam collocation points assigned to each region is chosen in
	proportion to its physical length. The largest-remainder procedure in
	\cref{eq:regional_counts} therefore gives \(125/208/167\) points for the
	reported partition.
	
	Within each region, samples are drawn independently from a uniform
	distribution. The first wall-region sample is replaced by \(\eta=0\), and
	the final tail-region sample is replaced by \(\eta=L\).
	
	An initial stratified batch is generated before the Adam loop. With the
	reported setting \texttt{resample\_interval}=1, this batch is used for the first Adam update and a newly sampled stratified batch is generated before
	every subsequent Adam update. No cumulative residual-adaptive
	point-enrichment procedure is used.
	
	
	\subsection{Adaptive regional residual balancing}
	\label{subsec:adaptive_weights}
	
	Regional losses may generate substantially different parameter-gradient
	magnitudes. The implementation therefore constructs adaptive coefficients
	from proxy gradient norms. The proxy parameter subset \(\theta_p\)
	consists of the weights and bias of the final neural-network layer together
	with the endpoint coefficient \(c_\theta\). For each region,
	
	\begin{equation}
		G_r
		=
		\left\lVert
		\nabla_{\theta_p}\mathcal{L}_r
		\right\rVert_2.
		\label{eq:gradient_norm}
	\end{equation}
	
	The Euclidean norm is formed jointly over the available proxy-parameter
	gradient components. A regional norm is considered valid only if it is
	finite and strictly greater than \(\varepsilon\). Invalid regional norms
	are replaced by the median of the valid regional norms; if none of the
	three norms is valid, the replacement value is set to unity.
	
	At the first adaptive update,
	
	\begin{equation}
		\overline{G}_r^{(0)}
		=
		G_r^{(0)}.
	\end{equation}
	
	At subsequent updates,
	
	\begin{equation}
		\overline{G}_r^{(k)}
		=
		\rho
		\overline{G}_r^{(k-1)}
		+
		(1-\rho)
		G_r^{(k)},
		\qquad
		\rho=0.90.
		\label{eq:gradient_ema}
	\end{equation}
	
	The normalized inverse-gradient coefficients are
	
	\begin{equation}
		q_r^{(k)}
		=
		\frac{
			\left(
			\overline{G}_r^{(k)}
			+
			\varepsilon
			\right)^{-1}
		}{
			\displaystyle
			\sum_j
			\left(
			\overline{G}_j^{(k)}
			+
			\varepsilon
			\right)^{-1}
		},
		\qquad
		\varepsilon=10^{-12}.
		\label{eq:inverse_weight}
	\end{equation}
	
	The safeguarded adaptive coefficients are
	
	\begin{equation}
		\boxed{
			\lambda_r^{(k)}
			=
			\gamma q_r^{(k)}
			+
			(1-\gamma)
			\frac13,
			\qquad
			\gamma=0.80
		}.
		\label{eq:safeguarded_weight}
	\end{equation}
	
	Thus,
	
	\begin{equation}
		\lambda_r
		\geq
		\frac{1-\gamma}{3}
		=
		0.0667.
	\end{equation}
	
	The three coefficients are renormalized so that
	
	\begin{equation}
		\sum_r\lambda_r=1.
	\end{equation}
	
	The weighted regional loss is
	
	\begin{equation}
		\mathcal{L}_{\mathrm{reg}}
		=
		\lambda_{\mathrm{w}}
		\mathcal{L}_{\mathrm{w}}
		+
		\lambda_{\mathrm{m}}
		\mathcal{L}_{\mathrm{m}}
		+
		\lambda_{\mathrm{t}}
		\mathcal{L}_{\mathrm{t}}.
		\label{eq:weighted_regional_loss}
	\end{equation}
	
	The proposed Adam objective is
	
	\begin{equation}
		\boxed{
			\mathcal{L}_{\mathrm{Adam}}
			=
			\alpha
			\mathcal{L}_{\mathrm{g}}
			+
			(1-\alpha)
			\mathcal{L}_{\mathrm{reg}},
			\qquad
			\alpha=0.50
		}.
		\label{eq:adam_objective}
	\end{equation}
	Combining \cref{eq:global_loss,eq:weighted_regional_loss,eq:adam_objective},
	the Adam objective can equivalently be written as
	
	\begin{equation}
		\mathcal{L}_{\mathrm{Adam}}
		=
		\sum_{r\in\{\mathrm{w},\mathrm{m},\mathrm{t}\}}
		w_r^{\mathrm{eff}}\mathcal{L}_r,
		\label{eq:effective_adam_objective}
	\end{equation}
	
	where
	
	\begin{equation}
		\boxed{
			w_r^{\mathrm{eff}}
			=
			\alpha\omega_r
			+
			(1-\alpha)\lambda_r
		}.
		\label{eq:effective_regional_weight}
	\end{equation}
	
	Thus, \(\lambda_r\) represents only the adaptive-balancing contribution
	and should not be interpreted as the complete coefficient multiplying
	\(\mathcal{L}_r\). The physical contribution
	\(\alpha\omega_r\) remains present in the Adam objective.
	
	Adaptive regional weights are updated at
	
	\begin{equation}
		1,\ 51,\ 101,\ 151,\ldots.
	\end{equation}
	
	Adaptive regional coefficients are updated only during Adam optimization. During L-BFGS refinement the coefficients are frozen and the optimization closure backpropagates the physical global residual \(\mathcal{L}_{\mathrm{g}}\) exclusively.
	
	
	\subsection{Baseline objectives}
	\label{subsec:baseline_objectives}
	
	For the global hard-constrained baseline,
	
	\begin{equation}
		\mathcal{L}_{A}
		=
		\mathcal{L}_{\mathrm{g}}.
		\label{eq:global_baseline}
	\end{equation}
	
	For the equal-regional hard-constrained PINN,
	
	\begin{equation}
		\lambda_{\mathrm{w}}
		=
		\lambda_{\mathrm{m}}
		=
		\lambda_{\mathrm{t}}
		=
		\frac13,
	\end{equation}
	
	and
	
	\begin{equation}
		\mathcal{L}_{A}
		=
		\frac12
		\mathcal{L}_{\mathrm{g}}
		+
		\frac12
		\left[
		\frac{
			\mathcal{L}_{\mathrm{w}}
			+
			\mathcal{L}_{\mathrm{m}}
			+
			\mathcal{L}_{\mathrm{t}}
		}{3}
		\right].
		\label{eq:equal_baseline}
	\end{equation}
	
	The adaptive regional formulation uses
	\cref{eq:adam_objective}. The regional coefficients are not adaptive in either baseline. The global
	formulation uses \(\mathcal{L}_{\mathrm{g}}\) alone, whereas the
	equal-regional formulation keeps
	\(\lambda_{\mathrm{w}}=\lambda_{\mathrm{m}}=\lambda_{\mathrm{t}}=1/3\)
	throughout Adam optimization. All three formulations subsequently use the
	same physical-global-residual L-BFGS objective.
	
	\subsection{Adam checkpointing and deterministic global-residual L-BFGS refinement}
	\label{subsec:optimization}
	
	The first optimization phase consists of
	
	\begin{equation}
		N_{\mathrm{Adam}}=15000
	\end{equation}
	
	completed Adam updates with initial learning rate
	
	\begin{equation}
		\mu_0=10^{-3}.
	\end{equation}
	
	An initial stratified collocation batch is generated before optimization.
	With \texttt{resample\_interval}=1, a newly sampled batch is generated
	before every Adam update after the first.
	
	The global norm of the gradients of all trainable model parameters is
	clipped according to
	
	\begin{equation}
		\boxed{
			\left\lVert
			\nabla_{(\theta,c_\theta)}
			\mathcal{L}_{\mathrm{Adam}}
			\right\rVert_2
			\leq1
		}.
		\label{eq:gradient_clipping}
	\end{equation}
	
	Adaptive regional coefficients are updated at completed Adam steps
	
	\begin{equation}
		1,\ 51,\ 101,\ 151,\ldots.
	\end{equation}
	
	Checkpoint selection is based on a deterministic uniform monitoring grid
	containing
	
	\begin{equation}
		N_{\mathrm{mon}}=800
	\end{equation}
	
	points over \([0,L]\). The monitor is the physical global residual
	
	\begin{equation}
		J_{\mathrm{mon}}
		=
		\mathcal{L}_{\mathrm{g}}^{\mathrm{mon}}.
		\label{eq:monitor}
	\end{equation}
	
	It is evaluated after completed Adam step 1, after every 100 completed
	updates, and after the final Adam update. Thus, a monitor entry indexed by
	step \(k\) refers to the model after \(k\) completed Adam parameter
	updates.
	
	At each scheduled monitor evaluation, the monitor value is first supplied
	to a \texttt{ReduceLROnPlateau} scheduler with patience 20 monitor
	evaluations and reduction factor 0.5. Whenever
	
	\begin{equation}
		J_{\mathrm{mon}}^{(k)}
		<
		J_{\mathrm{mon}}^{\mathrm{best}},
	\end{equation}
	
	the current model state and regional-weight state are stored.
	
	When the scheduled monitor first satisfies
	
	\begin{equation}
		J_{\mathrm{mon}}
		\leq
		10^{-6},
		\label{eq:lr_threshold}
	\end{equation}
	
	a one-time fine-stage learning-rate reduction is then applied to the
	current learning rate:
	
	\begin{equation}
		\mu_{\mathrm{new}}
		=
		\max
		\left(
		0.2\mu_{\mathrm{old}},
		10^{-5}
		\right).
		\label{eq:lr_drop}
	\end{equation}
	
	For the reported adaptive benchmark runs, this produces
	
	\begin{equation}
		10^{-3}
		\rightarrow
		2\times10^{-4}.
	\end{equation}
	
	The selected case first triggers this reduction at the scheduled monitor
	after completed Adam step 2000, the Blasius case at step 4900, and the
	Pohlhausen case at step 6000.
	
	After all Adam updates have been completed, the checkpoint having the
	smallest scheduled physical-global monitor is restored. For ARRB, the
	adaptive-weight state associated with that checkpoint is restored together
	with the model parameters. The complete adaptive-weight trajectory is
	retained separately for diagnostic plotting.
	
	Two deterministic L-BFGS stages are then applied. For stage \(j\),
	
	\begin{equation}
		N_{\mathrm{LBFGS},j}
		=
		\max
		\left(
		2000j,
		4N_c
		\right).
		\label{eq:lbfgs_grid}
	\end{equation}
	
	With \(N_c=500\),
	
	\begin{equation}
		N_{\mathrm{LBFGS},1}=2000,
		\qquad
		N_{\mathrm{LBFGS},2}=4000.
	\end{equation}
	
	The L-BFGS optimization closure minimizes only
	
	\begin{equation}
		\boxed{
			\min_{\theta,c_\theta}
			\mathcal{L}_{\mathrm{g}}
		}.
		\label{eq:lbfgs_objective}
	\end{equation}
	
	Thus, ARRB is an Adam-stage residual-conditioning mechanism; the
	quasi-Newton refinement objective is common to all three residual
	formulations.
	
	The PyTorch L-BFGS optimizer uses an initial step-size parameter of 0.5,
	at most 500 internal iterations per stage, and at most 1000 function
	evaluations per stage. The history size is 50, the gradient tolerance is
	\(10^{-12}\), the parameter-change tolerance is \(10^{-16}\), and the
	line search is strong Wolfe \cite{nocedal2006}.
	
	Optimization within L-BFGS stage \(j\) is performed on its corresponding
	2000- or 4000-point uniform grid. Stage acceptance, however, is determined
	using the separate 800-point deterministic monitor used for Adam
	checkpointing. Let \(J_{\mathrm{before}}\) and \(J_{\mathrm{after}}\)
	denote the monitor values immediately before and after a candidate
	L-BFGS stage. The stage is retained only if
	
	\begin{equation}
		J_{\mathrm{after}}
		\leq
		J_{\mathrm{before}}
		\label{eq:rollback}
	\end{equation}
	
	and \(J_{\mathrm{after}}\) is finite. Otherwise, the pre-stage model is
	restored.
	
	The overall sequence is therefore
	
	\begin{equation}
		\boxed{
			\text{Adam}
			\rightarrow
			\text{restore best monitored checkpoint}
			\rightarrow
			\text{L-BFGS}(N=2000)
			\rightarrow
			\text{L-BFGS}(N=4000)
		}.
		\label{eq:optimization_sequence}
	\end{equation} 

	\section{Computational algorithm}
	\label{sec:algorithm}
	
	\begin{algorithm}[H]
		\footnotesize
		\caption{Hard-constrained PINN with adaptive regional residual balancing
			for the generalized Falkner--Skan problem}
		\label{alg:hspinn_falkner_skan}
		
		\begin{algorithmic}[1]
			
			\Require
			\(\beta_0,\beta_1,L\), regional interfaces, network \(N_\theta\),
			\(N_A,N_c,\rho,\varepsilon,\gamma,\alpha\)
			
			\Ensure
			\(\hat f_\theta\), \(\hat f_\theta'\), \(\hat f_\theta''\),
			and \(\hat f_\theta''(0)\)
			
			\State Initialize neural parameters, set \(c_\theta\gets0\), and set
			\(\lambda_{\mathrm{w}}=\lambda_{\mathrm{m}}=\lambda_{\mathrm{t}}=1/3\)
			
			\State Define \(s=\eta/L\), \(x=2s-1\), and
			\(\hat f_\theta=b(\eta)+c_\theta H(s)+G(s)N_\theta(x)\)
			
			\State Construct the fixed 800-point monitoring grid
			
			\State Generate the initial stratified Adam collocation batch including
			\(\eta=0\) and \(\eta=L\)
			
			\Statex
			\State \textbf{Phase I: Adam optimization}
			
			\For{\(k=1,\ldots,N_A\)}
			
			\If{\(k>1\)}
			\State Generate a fresh stratified collocation batch
			\EndIf
			
			\State Compute
			\(\hat f_\theta'\), \(\hat f_\theta''\), \(\hat f_\theta'''\)
			and the governing-equation residual \(R_\theta\)
			
			\State Compute
			\(\mathcal{L}_{\mathrm{w}},
			\mathcal{L}_{\mathrm{m}},
			\mathcal{L}_{\mathrm{t}}\)
			and
			\(\mathcal{L}_{\mathrm{g}}
			=\sum_r\omega_r\mathcal{L}_r\)
			
			\If{\(k=1\) \textbf{or} \((k-1)\bmod50=0\)}
			
			\State Compute
			\(G_r=\|\nabla_{\theta_p}\mathcal{L}_r\|_2\)
			for \(r\in\{\mathrm{w},\mathrm{m},\mathrm{t}\}\)
			
			\State Replace invalid \(G_r\), update
			\(\overline{G}_r\), and compute
			\(q_r\propto(\overline{G}_r+\varepsilon)^{-1}\)
			
			\State Set
			\(\lambda_r=\gamma q_r+(1-\gamma)/3\)
			and normalize \(\sum_r\lambda_r=1\)
			
			\EndIf
			
			\State Compute
			\(\mathcal{L}_{\mathrm{reg}}=\sum_r\lambda_r\mathcal{L}_r\)
			and
			\(\mathcal{L}_{\mathrm{Adam}}
			=\alpha\mathcal{L}_{\mathrm{g}}
			+(1-\alpha)\mathcal{L}_{\mathrm{reg}}\)
			
			\State Backpropagate, clip the global model-gradient norm to \(1\),
			and perform one Adam update
			
			\If{\(k=1\) \textbf{or} \(k\bmod100=0\) \textbf{or} \(k=N_A\)}
			
			\State Evaluate the fixed-grid \(J_{\mathrm{mon}}\)
			
			\State Supply \(J_{\mathrm{mon}}\) to the plateau scheduler
			
			\If{\(J_{\mathrm{mon}}<J_{\mathrm{mon}}^{\mathrm{best}}\)}
			\State Save the model state and regional-weight state
			\EndIf
			
			\If{\(J_{\mathrm{mon}}\le10^{-6}\) for the first scheduled time}
			\State Apply the one-time learning-rate reduction
			\EndIf
			
			\EndIf
			\EndFor
			
			\State Restore the model and weight state associated with the best Adam monitor
			
			\Statex
			\State \textbf{Phase II: deterministic L-BFGS refinement}
			
			\For{\(j=1,2\)}
			
			\State Set
			\(N_{\mathrm{LBFGS},j}=\max(2000j,4N_c)\)
			and construct the L-BFGS uniform grid
			
			\State Evaluate and store \(J_{\mathrm{before}}\) on the independent
			800-point monitor
			
			\State Save the current model state
			
			\State Minimize \(\mathcal{L}_{\mathrm{g}}\) on the L-BFGS grid
			
			\State Evaluate \(J_{\mathrm{after}}\) on the 800-point monitor
			
			\If{\(J_{\mathrm{after}}\) is non-finite
				\textbf{or}
				\(J_{\mathrm{after}}>J_{\mathrm{before}}\)}
			\State Restore the pre-stage model
			\EndIf
			
			\EndFor
			
			\State Evaluate the accepted solution on 5000 uniform points and compute
			wall-shear, residual, and reference-error metrics
			
			\State \Return final solution and validation metrics
			
		\end{algorithmic}
	\end{algorithm}

	
	\section{Reference solutions and validation metrics}
	\label{sec:validation}
	
	\subsection{Numerical boundary-value reference}
	\label{subsec:bvp_reference}
	
	For quantitative validation, a numerical reference is computed using the
	SciPy boundary-value solver \cite{virtanen2020}. Introduce
	
	\begin{equation}
		y_1=f,
		\qquad
		y_2=f',
		\qquad
		y_3=f''.
	\end{equation}
	
	Then
	
	\begin{equation}
		y_1'=y_2,
	\end{equation}
	
	\begin{equation}
		y_2'=y_3,
	\end{equation}
	
	and
	
	\begin{equation}
		y_3'
		=
		-\beta_0y_1y_3
		-
		\beta_1
		\left(
		1-y_2^2
		\right).
		\label{eq:bvp_system}
	\end{equation}
	
	The boundary conditions are
	
	\begin{equation}
		y_1(0)=0,
		\qquad
		y_2(0)=0,
		\qquad
		y_2(L)=1.
	\end{equation}
	
	The same truncated domain \(L=12\) is used for the PINN and numerical BVP
	reference.
	
	The numerical solver begins from a uniform mesh containing 1200 points and
	uses residual tolerance
	
	\begin{equation}
		10^{-10}.
	\end{equation}
	
	The maximum permitted number of mesh nodes is
	
	\begin{equation}
		30000.
	\end{equation}
	
	The initial BVP velocity profile is based on
	
	\begin{equation}
		f'(\eta)
		\approx
		1-e^{-\eta},
	\end{equation}
	
	with
	
	\begin{equation}
		f(\eta)
		\approx
		\eta-1+e^{-\eta}.
	\end{equation}
	
	Several candidate wall-shear initial guesses are attempted if necessary
	until the BVP solver converges.
	
	The resulting BVP solution is treated as a high-accuracy numerical
	reference rather than an exact solution. The specified solver tolerance
	controls the convergence criterion of the numerical algorithm and should
	not be interpreted by itself as a rigorous absolute error bound for every
	reported solution quantity.
	
	
	\subsection{Analytical Pohlhausen benchmark}
	\label{subsec:pohlhausen}
	
	For
	
	\begin{equation}
		\beta_0=0,
		\qquad
		\beta_1=1,
	\end{equation}
	
	the governing equation becomes
	
	\begin{equation}
		f'''
		+
		1
		-
		(f')^2
		=
		0.
	\end{equation}
	
	Define
	
	\begin{equation}
		a
		=
		\operatorname{arctanh}
		\sqrt{\frac23},
	\end{equation}
	
	and
	
	\begin{equation}
		\xi
		=
		\frac{\eta}{\sqrt2}
		+
		a.
	\end{equation}
	
	The analytical semi-infinite-domain solution is
	
	\begin{equation}
		f'(\eta)
		=
		3\tanh^2(\xi)-2,
		\label{eq:pohlhausen_fp}
	\end{equation}
	
	\begin{equation}
		f''(\eta)
		=
		3\sqrt2
		\tanh(\xi)
		\operatorname{sech}^2(\xi),
		\label{eq:pohlhausen_fpp}
	\end{equation}
	
	and
	
	\begin{equation}
		f(\eta)
		=
		\eta
		-
		3\sqrt2
		\left[
		\tanh(\xi)-\tanh(a)
		\right].
		\label{eq:pohlhausen_f}
	\end{equation}
	
	The third derivative is
	
	\begin{equation}
		f'''(\eta)
		=
		3
		\operatorname{sech}^2(\xi)
		\left[
		1-3\tanh^2(\xi)
		\right].
	\end{equation}
	
	The exact wall shear is
	
	\begin{equation}
		\boxed{
			f''(0)
			=
			\frac{2}{\sqrt3}
			=
			1.154700538379\ldots
		}.
		\label{eq:pohlhausen_wall}
	\end{equation}
	
	The analytical solution satisfies
	
	\begin{equation}
		f'(\eta)\rightarrow1
		\qquad
		\text{as}
		\qquad
		\eta\rightarrow\infty,
	\end{equation}
	
	whereas the PINN and numerical BVP calculations impose \(f'(12)=1\)
	exactly. The analytical semi-infinite problem and the finite-domain problem
	therefore do not satisfy precisely the same terminal condition.
	
	Accordingly, \(2/\sqrt3\) is used as the independent analytical
	Pohlhausen wall-shear benchmark. The implementation additionally computes
	the finite-domain BVP wall shear as a truncation diagnostic, while all
	reported discrete profile errors are evaluated against the finite-domain
	BVP posed on the same interval as the PINN. This distinction separates the analytical wall-shear benchmark from the finite-domain reference used for profile comparison.
	
	
	\subsection{Validation metrics}
	\label{subsec:metrics}
	
	Final metrics are evaluated on a uniform grid containing
	
	\begin{equation}
		N_e=5000
	\end{equation}
	
	points.
	
	The residual MSE is
	
	\begin{equation}
		E_R^{\mathrm{MSE}}
		=
		\omega_{\mathrm{w}}
		\mathcal{L}_{\mathrm{w}}^{\mathrm{eval}}
		+
		\omega_{\mathrm{m}}
		\mathcal{L}_{\mathrm{m}}^{\mathrm{eval}}
		+
		\omega_{\mathrm{t}}
		\mathcal{L}_{\mathrm{t}}^{\mathrm{eval}}.
		\label{eq:eval_residual_mse}
	\end{equation}
	
	The maximum residual is
	
	\begin{equation}
		E_R^{\max}
		=
		\max_i
		|R_\theta(\eta_i)|.
		\label{eq:max_residual}
	\end{equation}
	
	The relative velocity-profile error is
	
	\begin{equation}
		E_{f'}^{L_2}
		=
		\frac{
			\left\lVert
			f'_{\mathrm{PINN}}
			-
			f'_{\mathrm{ref}}
			\right\rVert_2
		}{
			\left\lVert
			f'_{\mathrm{ref}}
			\right\rVert_2
		}.
		\label{eq:fp_l2}
	\end{equation}
	
	The relative shear-profile error is
	
	\begin{equation}
		E_{f''}^{L_2}
		=
		\frac{
			\left\lVert
			f''_{\mathrm{PINN}}
			-
			f''_{\mathrm{ref}}
			\right\rVert_2
		}{
			\left\lVert
			f''_{\mathrm{ref}}
			\right\rVert_2
		}.
		\label{eq:fpp_l2}
	\end{equation}
	
	The maximum velocity error is
	
	\begin{equation}
		E_{f'}^{\max}
		=
		\max_i
		\left|
		f'_{\mathrm{PINN}}(\eta_i)
		-
		f'_{\mathrm{ref}}(\eta_i)
		\right|.
	\end{equation}
	
	The wall-shear error is
	
	\begin{equation}
		E_{\mathrm{w}}
		=
		\left|
		f''_{\mathrm{PINN}}(0)
		-
		f''_{\mathrm{benchmark}}(0)
		\right|.
		\label{eq:wall_error}
	\end{equation}
	
	For Blasius and the selected case, the benchmark is the numerical BVP wall
	shear. For Pohlhausen, the benchmark is \(2/\sqrt3\).
	
	
	\section{Computational configuration}
	\label{sec:configuration}
	
	The principal settings used for the reported calculations are summarized in
	\cref{tab:configuration}.
	
	\begin{table}[htbp]
		
		\centering
		\small
		\renewcommand{\arraystretch}{0.94}
		
		\caption{
			Main computational parameters of the proposed hard-constrained PINN with
			ARRB.
		}
		
		\label{tab:configuration}
		
		\resizebox{0.96\textwidth}{!}{%
			\begin{tabular}{ll}
				
				\toprule
				Parameter & Value \\
				\midrule
				
				Truncated computational domain
				&
				\(\eta\in[0,12]\)
				\\
				
				Network architecture
				&
				\([1,100,100,1]\)
				\\
				
				Activation function
				&
				tanh
				\\
				
				Numerical precision
				&
				64-bit floating point
				\\
				
				Publication computation device
				&
				CPU
				\\
				
				Adam collocation points
				&
				500
				\\
				
				Wall/middle/tail counts
				&
				125 / 208 / 167
				\\
				
				Regional intervals
				&
				\([0,3]\), \((3,8)\), \([8,12]\)
				\\
				
				Physical global weights
				&
				\(1/4,\ 5/12,\ 1/3\)
				\\
				
				Residual-adaptive point refinement
				&
				not used
				\\
				
				Adam steps
				&
				15000
				\\
				
				Initial Adam learning rate
				&
				\(1\times10^{-3}\)
				\\
				
				Threshold-triggered LR factor
				&
				0.20
				\\
				
				Threshold-triggered LR floor
				&
				\(1\times10^{-5}\)
				\\
				
				Fine-stage threshold
				&
				\(1\times10^{-6}\)
				\\
				
				Plateau scheduler factor
				&
				0.5
				\\
				
				Plateau scheduler patience
				&
				20 monitor evaluations
				\\
				
				Adaptive-weight updates during Adam
				&
				steps \(1,51,101,\ldots\)
				\\
				
				EMA factor
				&
				\(\rho=0.90\)
				\\
				
				Gradient-norm stabilizer
				&
				\(\varepsilon=10^{-12}\)
				\\
				
				Adaptive blend
				&
				\(\gamma=0.80\)
				\\
				
				Global-loss fraction
				&
				\(\alpha=0.50\)
				\\
				
				Gradient clipping
				&
				global model-parameter norm \(\leq1\)
				\\
				
				Fixed monitoring points
				&
				800
				\\
				
				Monitoring schedule
				&
				after step 1, every 100 completed updates, and after final update
				\\
				
				L-BFGS stages
				&
				2
				\\
				
				L-BFGS optimization grids
				&
				2000 and 4000 points
				\\
				
				L-BFGS initial step-size parameter
				&
				0.5
				\\
				
				Maximum L-BFGS iterations
				&
				500 per stage
				\\
				
				Maximum L-BFGS evaluations
				&
				1000 per stage
				\\
				
				L-BFGS history size
				&
				50
				\\
				
				L-BFGS gradient tolerance
				&
				\(10^{-12}\)
				\\
				
				L-BFGS change tolerance
				&
				\(10^{-16}\)
				\\
				
				L-BFGS line search
				&
				strong Wolfe
				\\
				L-BFGS acceptance monitor
				&
				same independent 800-point fixed grid
				\\
				Final evaluation points
				&
				5000
				\\
				
				Numerical BVP tolerance
				&
				\(10^{-10}\)
				\\
				
				Numerical BVP initial mesh
				&
				1200 points
				\\
				
				Maximum BVP mesh nodes
				&
				30000
				\\
				
				Random seed
				&
				42
				\\
				
				\bottomrule
				
			\end{tabular}%
		}
		
	\end{table}
	
	
	\section{Numerical results}
	\label{sec:results}
	
	All results reported in this section use the physical global residual
	defined in \cref{eq:global_loss}. The same global residual is used for
	fixed-grid monitoring, L-BFGS refinement, and final residual evaluation.
	
	
	\subsection{Accuracy for the three benchmark cases}
	\label{subsec:case_results}
	
	The final 5000-point evaluation-grid results are summarized in
	\cref{tab:case_results}.
	
	\begin{table}[htbp]
		
		\centering
		\small
		
		\caption{
			Evaluation-grid accuracy of the hard-constrained PINN with adaptive regional
			residual balancing.
		}
		
		\label{tab:case_results}
		
		\resizebox{\textwidth}{!}{%
			\begin{tabular}{lcccccc}
				
				\toprule
				
				Case
				&
				\((\beta_0,\beta_1)\)
				&
				PINN \(f''(0)\)
				&
				Reference \(f''(0)\)
				&
				Wall error
				&
				Residual MSE
				&
				Relative \(L_2(f'')\)
				\\
				
				\midrule
				
				Blasius
				&
				\((0.50,0.00)\)
				&
				0.3320572256
				&
				0.3320573362
				&
				\(1.106\times10^{-7}\)
				&
				\(5.039\times10^{-8}\)
				&
				\(3.586\times10^{-4}\)
				\\
				
				Selected
				&
				\((0.75,0.50)\)
				&
				0.8997161394
				&
				0.8997168085
				&
				\(6.691\times10^{-7}\)
				&
				\(6.357\times10^{-9}\)
				&
				\(4.188\times10^{-5}\)
				\\
				
				Pohlhausen
				&
				\((0.00,1.00)\)
				&
				1.1546922871
				&
				1.1547005384
				&
				\(8.251\times10^{-6}\)
				&
				\(6.483\times10^{-8}\)
				&
				\(1.735\times10^{-4}\)
				\\
				
				\bottomrule
				
			\end{tabular}%
		}
		
		\vspace{0.4em}
		
		\parbox{0.98\textwidth}{\footnotesize
			\emph{Note:} For Pohlhausen, the wall-shear error is computed against the
			analytical semi-infinite-domain value \(2/\sqrt3\), while the profile
			errors are computed against the finite-domain BVP reference.}
		
	\end{table}
	
	The relative \(L_2\) velocity errors for the Blasius, selected, and
	Pohlhausen cases are
	
	\begin{equation}
		2.093\times10^{-5},
		\qquad
		2.848\times10^{-6},
		\qquad
		1.960\times10^{-5},
	\end{equation}
	
	respectively.
	
	Their maximum absolute velocity errors are
	
	\begin{equation}
		4.121\times10^{-5},
		\qquad
		6.835\times10^{-6},
		\qquad
		4.332\times10^{-5},
	\end{equation}
	
	and the corresponding maximum absolute equation residuals are
	
	\begin{equation}
		1.410\times10^{-3},
		\qquad
		7.540\times10^{-4},
		\qquad
		3.406\times10^{-3}.
	\end{equation}
	
	For all three cases, the hard-boundary diagnostics give
	
	\begin{equation}
		|\hat f_\theta(0)|
		=
		|\hat f_\theta'(0)|
		=
		|\hat f_\theta'(L)-1|
		=
		0
	\end{equation}
	
	to the displayed numerical precision.
	
	The adaptive coefficients restored with the best Adam checkpoints are
	
	\begin{align}
		\text{Blasius:}\quad
		&
		(0.7044,\ 0.2130,\ 0.0826),
		\\
		\text{Selected:}\quad
		&
		(0.7276,\ 0.1816,\ 0.0908),
		\\
		\text{Pohlhausen:}\quad
		&
		(0.4552,\ 0.3177,\ 0.2271).
	\end{align}
	These coefficients are not updated during the subsequent L-BFGS stages. The complete 15000-step adaptive-coefficient trajectory is retained separately for diagnostic plotting even when the best Adam checkpoint occurs before the final Adam update.	
	
	\begin{figure}[htbp]
		
		\centering
		
		\begin{subfigure}[t]{0.49\textwidth}
			\centering
			\safeincludegraphics[width=\linewidth]
			{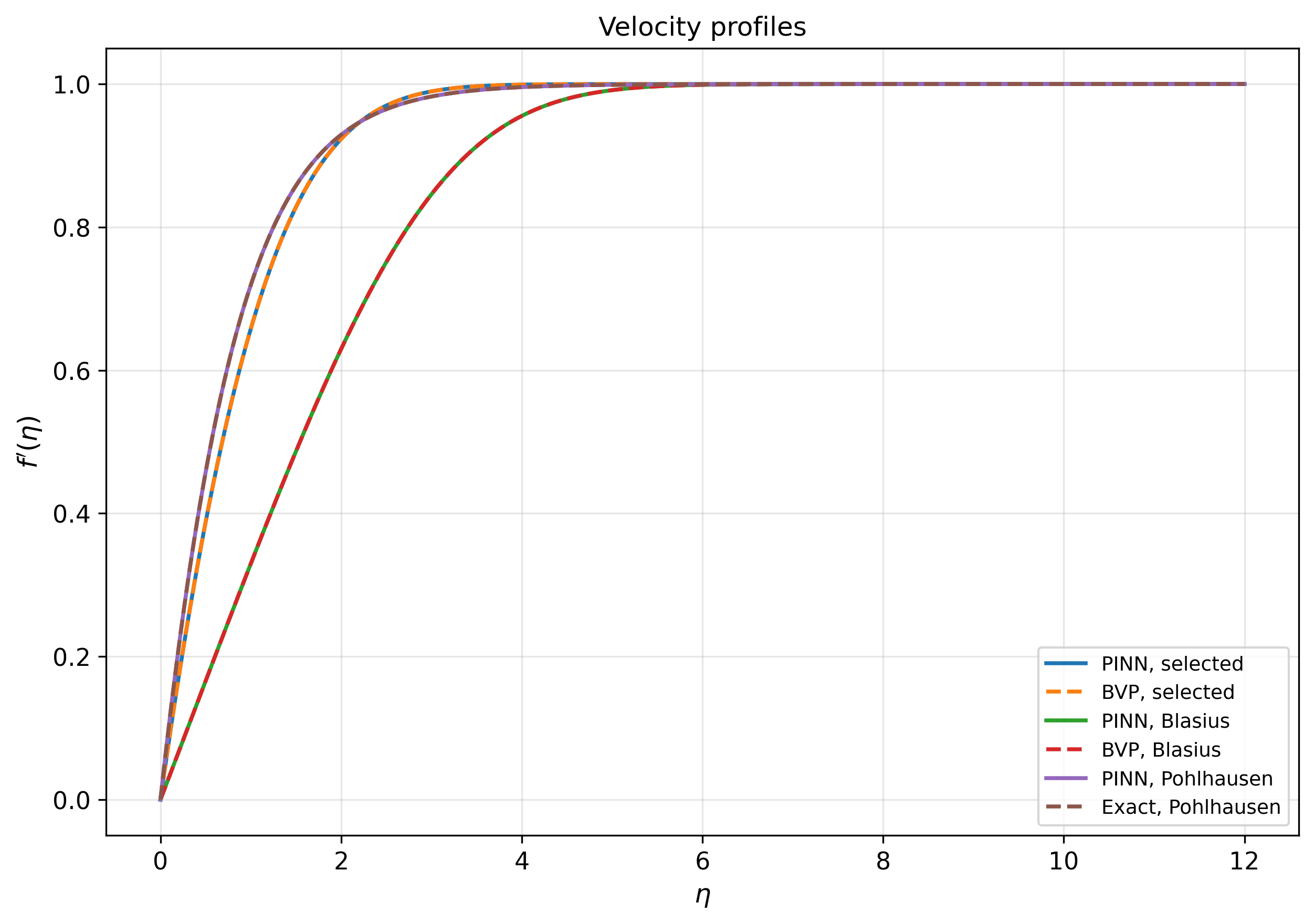}
			{Velocity-profile comparison}
			\caption{Velocity profiles.}
			\label{fig:velocity_profiles}
		\end{subfigure}
		\hfill
		\begin{subfigure}[t]{0.49\textwidth}
			\centering
			\safeincludegraphics[width=\linewidth]
			{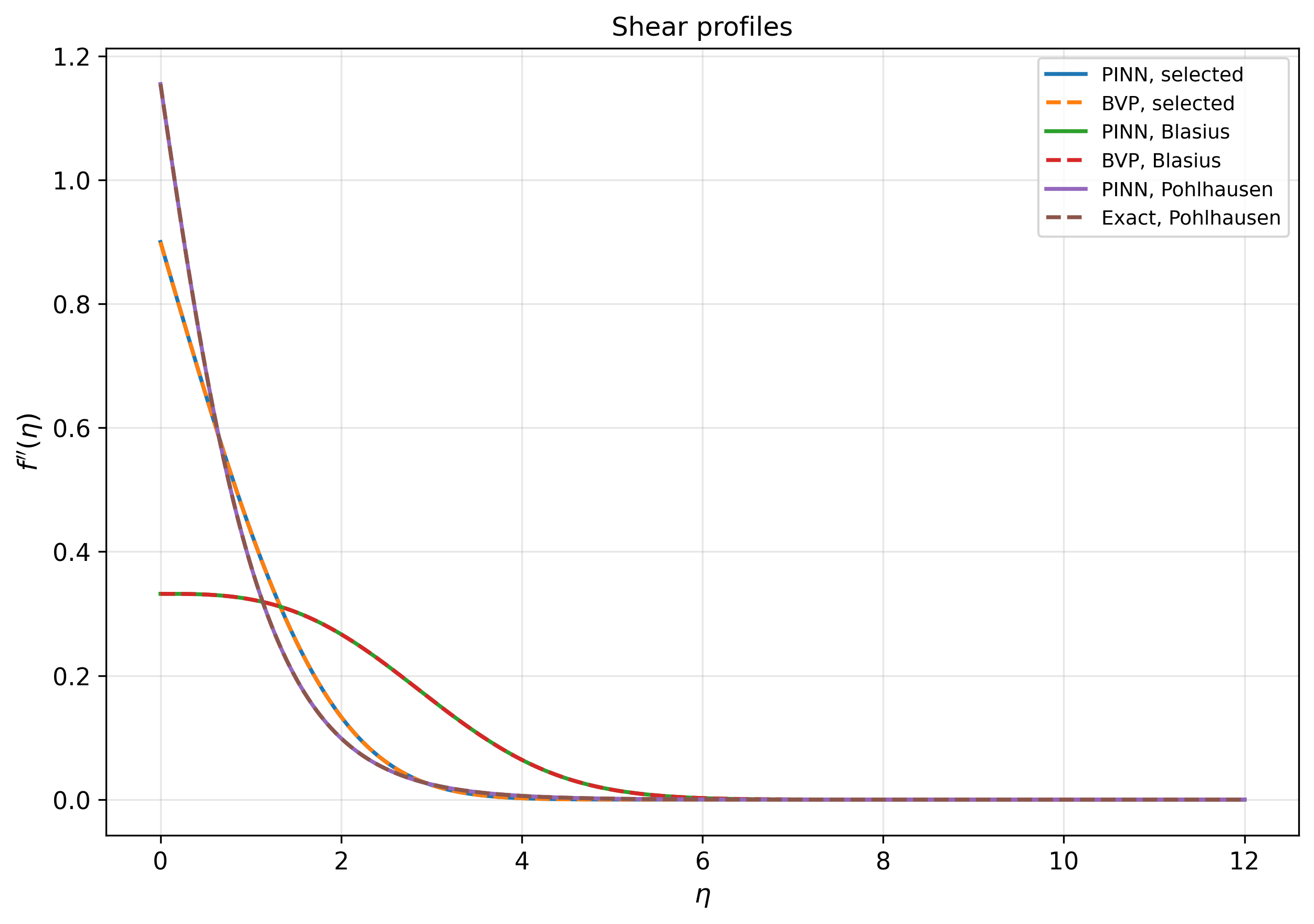}
			{Shear-profile comparison}
			\caption{Shear profiles.}
			\label{fig:shear_profiles}
		\end{subfigure}
		
		\caption{
			PINN and reference profiles for the selected Falkner--Skan, Blasius, and
			Pohlhausen cases: (a) normalized velocity \(f'(\eta)\); (b) shear profile
			\(f''(\eta)\).
		}
		
		\label{fig:profile_comparison}
		
	\end{figure}
	
	As shown in \cref{fig:profile_comparison}, the predicted velocity and
	shear profiles closely follow the numerical reference solutions for all
	three benchmark cases.
	
	Relative to the PINN Blasius wall shear, the selected Falkner--Skan case
	increases \(f''(0)\) by
	
	\begin{equation}
		0.5676589138,
	\end{equation}
	
	or \(170.952\%\).
	
	The Pohlhausen increase relative to Blasius is
	
	\begin{equation}
		0.8226350615,
	\end{equation}
	
	or \(247.739\%\).
	
	
	\begin{figure}[htbp]
		
		\centering
		
		\safeincludegraphics[width=0.78\textwidth]
		{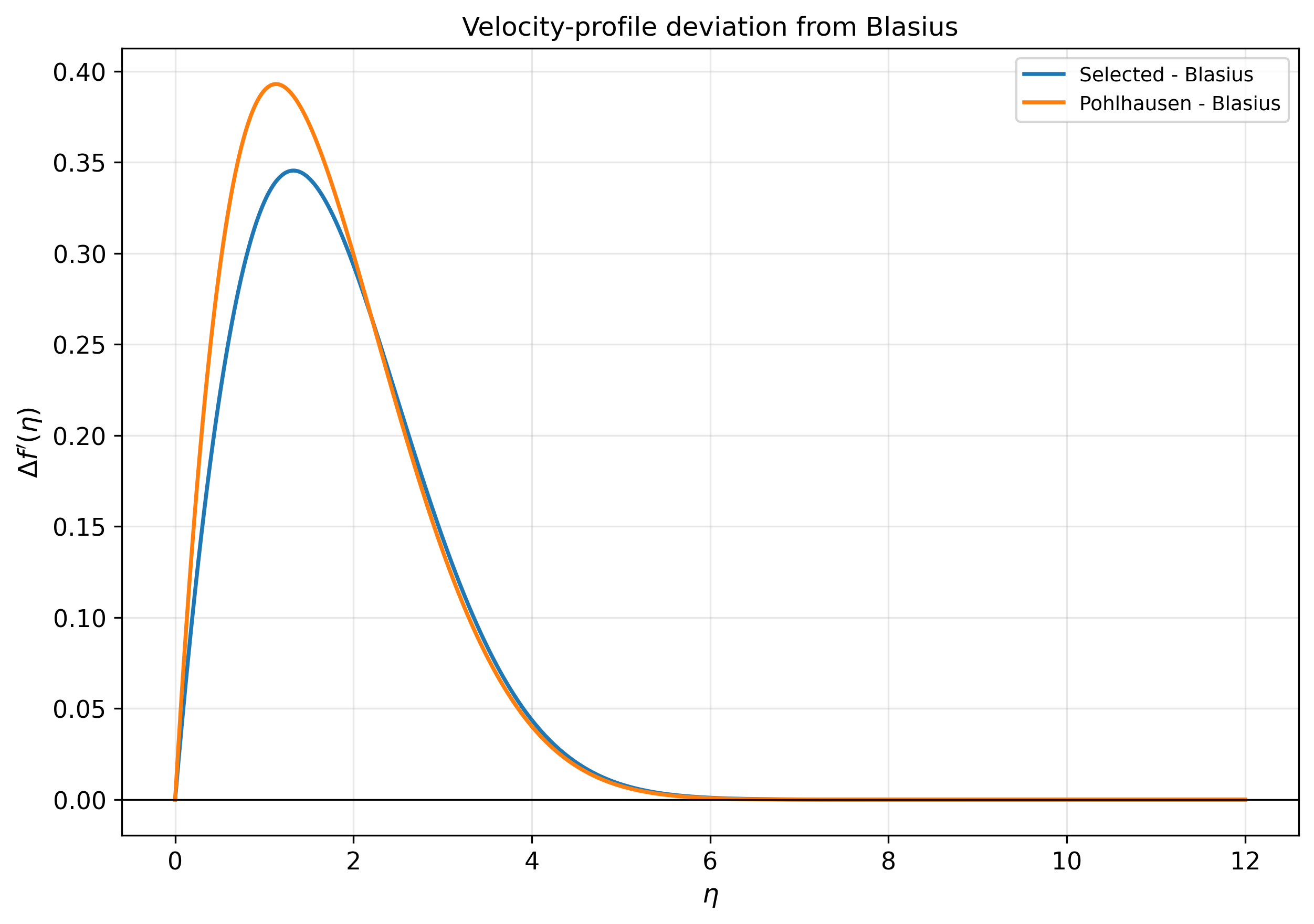}
		{Velocity-profile deviation from Blasius}
		
		\caption{
			Reference velocity-profile deviations from the Blasius solution for the
			selected favorable-pressure-gradient Falkner--Skan case and the Pohlhausen
			benchmark.
		}
		
		\label{fig:velocity_deviation}
		
	\end{figure}
	
	The corresponding changes in the velocity profiles are illustrated in
	\cref{fig:velocity_deviation}.
	
	
	\subsection{Selected-case diagnostics}
	\label{subsec:selected_results}
	
	For the selected case,
	
	\begin{equation}
		(\beta_0,\beta_1)
		=
		(0.75,0.50),
	\end{equation}
	
	the final wall shear is
	
	\begin{equation}
		\hat f_\theta''(0)
		=
		0.8997161394,
	\end{equation}
	
	while the numerical BVP reference is
	
	\begin{equation}
		f_{\mathrm{ref}}''(0)
		=
		0.8997168085.
	\end{equation}
	
	The resulting wall-shear error is
	
	\begin{equation}
		E_{\mathrm{w}}
		=
		6.691\times10^{-7}.
	\end{equation}
	
	The final 5000-point evaluation-grid physical residual MSE is
	
	\begin{equation}
		E_R^{\mathrm{MSE}}
		=
		6.357\times10^{-9},
	\end{equation}
	
	and the maximum absolute residual is
	
	\begin{equation}
		E_R^{\max}
		=
		7.540\times10^{-4}.
	\end{equation}
	
	The relative profile errors are
	
	\begin{equation}
		E_{f'}^{L_2}
		=
		2.848\times10^{-6},
		\qquad
		E_{f''}^{L_2}
		=
		4.188\times10^{-5}.
	\end{equation}
	
	The maximum absolute velocity-profile error is
	
	\begin{equation}
		E_{f'}^{\max}
		=
		6.835\times10^{-6}.
	\end{equation}
	
	The selected-case regional residual MSEs are
	
	\begin{equation}
		\mathcal{L}_{\mathrm{w}}
		=
		1.234\times10^{-8},
	\end{equation}
	
	\begin{equation}
		\mathcal{L}_{\mathrm{m}}
		=
		1.323\times10^{-9},
	\end{equation}
	
	and
	
	\begin{equation}
		\mathcal{L}_{\mathrm{t}}
		=
		8.161\times10^{-9}.
	\end{equation}
	
	The adaptive coefficients restored with the selected-case best Adam
	checkpoint are
	
	\begin{equation}
		\boxed{
			(
			\lambda_{\mathrm{w}},
			\lambda_{\mathrm{m}},
			\lambda_{\mathrm{t}}
			)
			=
			(0.7276,\ 0.1816,\ 0.0908)
		}.
	\end{equation}
	
	Because \(\alpha=0.50\), the corresponding complete coefficients
	multiplying the three regional losses during the Adam objective at this
	checkpoint are, from \cref{eq:effective_regional_weight},
	
	\begin{equation}
		\boxed{
			(
			w_{\mathrm{w}}^{\mathrm{eff}},
			w_{\mathrm{m}}^{\mathrm{eff}},
			w_{\mathrm{t}}^{\mathrm{eff}}
			)
			\approx
			(0.4888,\ 0.2991,\ 0.2121)
		}.
	\end{equation}
	
	Thus, the adaptive coefficients \(\lambda_r\) should not be interpreted
	as the complete regional weights of the Adam objective.
	
	\begin{figure}[htbp]
		
		\centering
		
		\begin{subfigure}[t]{0.49\textwidth}
			\centering
			\safeincludegraphics[width=\linewidth]
			{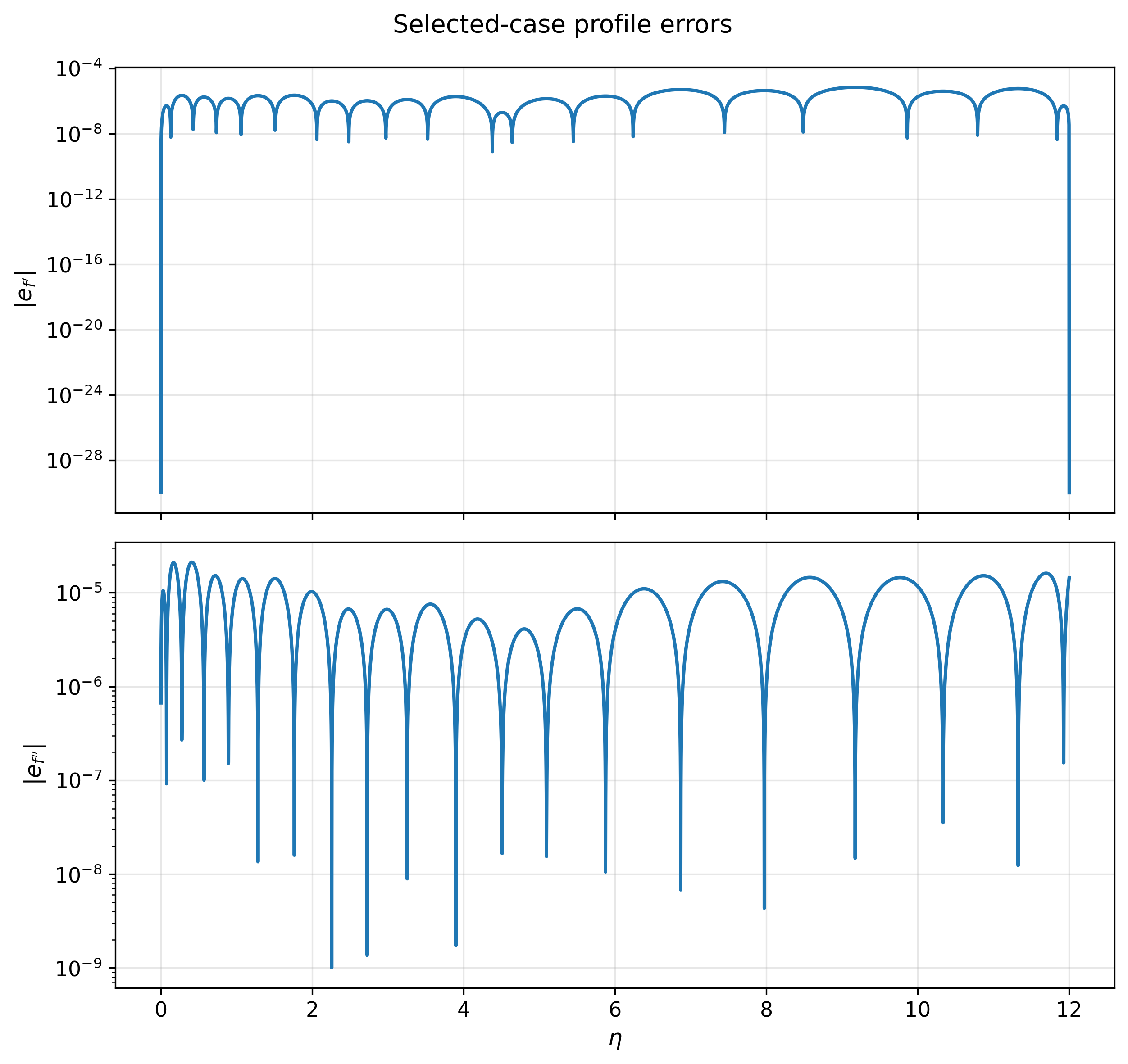}
			{Selected-case profile errors}
			\caption{Pointwise profile errors.}
			\label{fig:selected_error}
		\end{subfigure}
		\hfill
		\begin{subfigure}[t]{0.49\textwidth}
			\centering
			\safeincludegraphics[width=\linewidth]
			{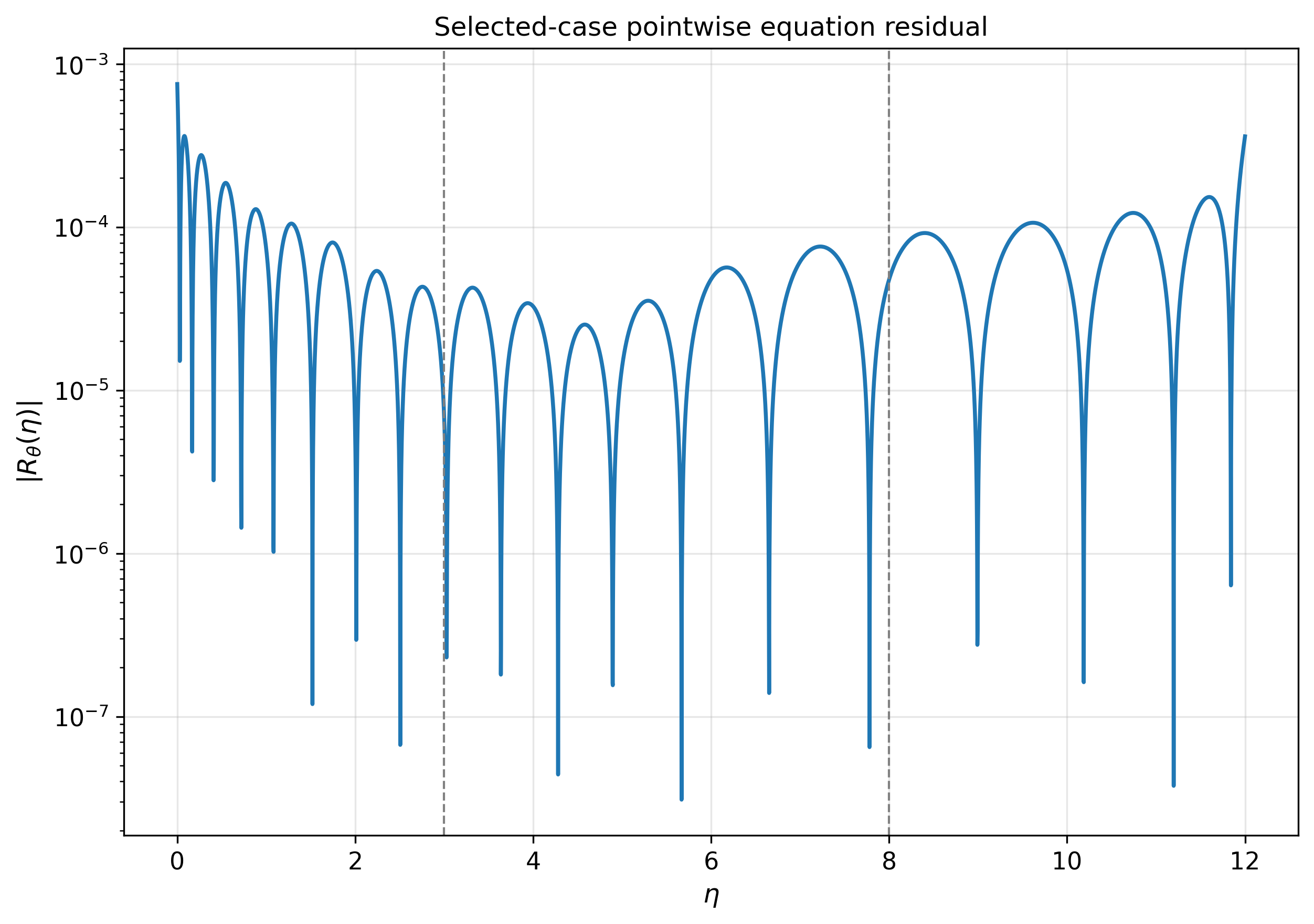}
			{Selected-case equation residual}
			\caption{Pointwise residual magnitude.}
			\label{fig:selected_residual}
		\end{subfigure}
		
		\caption{
			Pointwise diagnostics for the selected
			\((\beta_0,\beta_1)=(0.75,0.50)\) case.
		}
		
		\label{fig:selected_diagnostics}
		
	\end{figure}
	
	The selected case first satisfies the \(10^{-6}\) threshold at the scheduled fixed-monitor evaluation after completed Adam step 2000. The best Adam monitor is
	
	\begin{equation}
		6.997\times10^{-9}.
	\end{equation}
	
	The first L-BFGS stage changes this value to
	
	\begin{equation}
		6.801\times10^{-9},
	\end{equation}
	
	and the second stage further reduces it to
	
	\begin{equation}
		6.798\times10^{-9}.
	\end{equation}
	
	Both stages are therefore accepted.
	
	
	\begin{figure}[htbp]
		
		\centering
		
		\safeincludegraphics[width=0.82\textwidth]
		{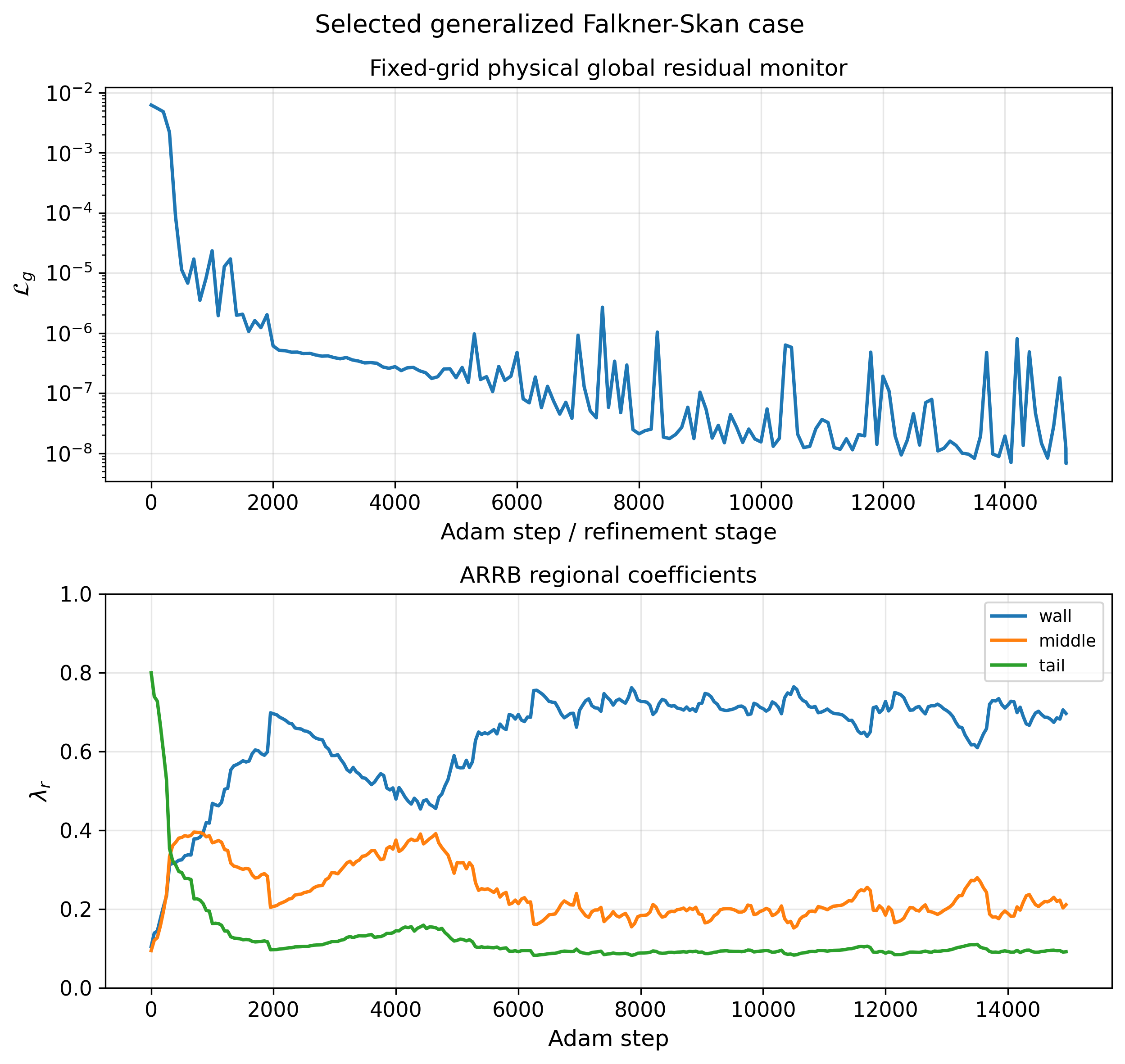}
		{Fixed-grid monitor and adaptive regional weights}
		
		\caption{Deterministic fixed-grid physical-global-residual monitor and complete
			Adam-stage adaptive-coefficient trajectory for the selected
			\((\beta_0,\beta_1)=(0.75,0.50)\) case. The coefficients used for reporting
			the restored model are those associated with the best Adam checkpoint,
			whereas the plotted coefficient history retains all adaptive updates for
			diagnostic purposes.}
		
		\label{fig:monitor}
		
	\end{figure}
	
	\Cref{fig:monitor} shows the deterministic fixed-grid convergence history
	together with the evolution of the adaptive regional coefficients.
	
	
	\subsection{Regional residual distribution}
	\label{subsec:regional_results}
	
	\Cref{tab:regional_results} summarizes the final evaluation-grid regional
	residual MSEs together with the adaptive coefficients restored with the
	best Adam checkpoints.
	
	\begin{table}[htbp]
		
		\centering
		\small
		
		\caption{
			Evaluation-grid regional residual MSEs and adaptive coefficients restored
			with the best Adam checkpoints.
		}
		
		\label{tab:regional_results}
		
		\resizebox{\textwidth}{!}{%
			\begin{tabular}{lcccccc}
				
				\toprule
				Case
				&
				\(\mathcal{L}_{\mathrm{w}}\)
				&
				\(\mathcal{L}_{\mathrm{m}}\)
				&
				\(\mathcal{L}_{\mathrm{t}}\)
				&
				\(\lambda_{\mathrm{w}}\)
				&
				\(\lambda_{\mathrm{m}}\)
				&
				\(\lambda_{\mathrm{t}}\)
				\\
				
				\midrule
				
				Blasius
				&
				\(4.966\times10^{-8}\)
				&
				\(6.246\times10^{-8}\)
				&
				\(3.586\times10^{-8}\)
				&
				0.7044
				&
				0.2130
				&
				0.0826
				\\
				
				Selected
				&
				\(1.234\times10^{-8}\)
				&
				\(1.323\times10^{-9}\)
				&
				\(8.161\times10^{-9}\)
				&
				0.7276
				&
				0.1816
				&
				0.0908
				\\
				
				Pohlhausen
				&
				\(1.864\times10^{-7}\)
				&
				\(8.176\times10^{-9}\)
				&
				\(4.446\times10^{-8}\)
				&
				0.4552
				&
				0.3177
				&
				0.2271
				\\
				
				\bottomrule
				
			\end{tabular}%
		}
		
	\end{table}
	
	The adaptive coefficients are optimization-balancing quantities. A larger
	\(\lambda_r\) does not imply greater physical importance. Under the
	inverse-gradient construction, a larger coefficient indicates that the
	corresponding regional loss would otherwise generate a comparatively
	smaller gradient over the selected proxy parameter subset.
	
	Moreover, \(\lambda_r\) constitutes only the adaptive portion of the Adam
	objective. The complete instantaneous regional coefficient is
	\(w_r^{\mathrm{eff}}
	=\alpha\omega_r+(1-\alpha)\lambda_r\), so the physical region-length weighting remains present throughout Adam training. ------------------------------------------------------------
	
	\subsection{Residual-objective ablation}
	\label{subsec:ablation}
	
	The selected favorable-pressure-gradient case is used for a matched
	single-seed comparison of the global, equal-regional, and adaptive-regional
	objectives. Within the reported comparison, the three formulations use the
	same hard-constrained trial representation, architecture, initialization
	seed, stochastic sampling procedure, Adam settings, deterministic monitor,
	checkpoint rule, and L-BFGS refinement protocol. The intended
	methodological difference is restricted to the residual objective used
	during Adam: physical global, equal-regional mixture, or adaptively
	balanced regional mixture.
	
	Because the deterministic monitor is evaluated only at scheduled
	intervals, the convergence indicator is defined as the earliest scheduled
	Adam-monitor evaluation for which
	
	\begin{equation}
		J_{\mathrm{mon}}
		\leq
		10^{-6}.
		\label{eq:ablation_threshold}
	\end{equation}
	
	This quantity is therefore a first \emph{monitored} threshold-crossing
	step rather than the exact optimizer iteration at which the evolving
	network first satisfies the inequality.
	
	\begin{table}[htbp]
		
		\centering
		\small
		
		\caption{
			Matched single-seed selected-case residual-objective ablation for
			\((\beta_0,\beta_1)=(0.75,0.50)\).
		}
		
		\label{tab:ablation}
		
		\resizebox{\textwidth}{!}{%
			\begin{tabular}{lccccc}
				
				\toprule
				Method
				&
				Residual MSE
				&
				\(\max |R|\)
				&
				Wall-shear error
				&
				Relative \(L_2(f'')\)
				&
				First monitored threshold step
				\\
				\midrule
				
				Global hard-constrained PINN
				&
				\(1.347\times10^{-8}\)
				&
				\(1.368\times10^{-3}\)
				&
				\(2.947\times10^{-6}\)
				&
				\(6.056\times10^{-5}\)
				&
				1800
				\\
				
				Equal-regional hard-constrained PINN
				&
				\(1.277\times10^{-8}\)
				&
				\(1.185\times10^{-3}\)
				&
				\(2.422\times10^{-6}\)
				&
				\(5.842\times10^{-5}\)
				&
				1500
				\\
				
				ARRB hard-constrained PINN
				&
				\(6.357\times10^{-9}\)
				&
				\(7.540\times10^{-4}\)
				&
				\(6.691\times10^{-7}\)
				&
				\(4.188\times10^{-5}\)
				&
				2000
				\\
				
				\bottomrule
			\end{tabular}%
		}
		
		\vspace{0.4em}
		
		\parbox{0.98\textwidth}{\footnotesize
			\emph{Note:} Percentage reductions reported in the text are computed from
			the unrounded numerical values. ``First monitored threshold step'' denotes
			the earliest scheduled fixed-monitor evaluation satisfying
			\(J_{\mathrm{mon}}\le10^{-6}\).}
		
	\end{table}
	
	For this matched seed, ARRB reduces the final residual MSE by
	approximately \(52.81\%\) relative to the global hard-constrained
	formulation and by \(50.24\%\) relative to equal-regional weighting. The
	wall-shear error is reduced by approximately \(77.30\%\) relative to the
	global formulation and by \(72.37\%\) relative to equal-regional
	weighting.
	
	ARRB also gives the smallest maximum absolute residual and relative
	shear-profile error among the three objectives in this experiment. These
	differences concern final approximation accuracy rather than the speed at
	which the prescribed monitored threshold is reached.
	
	Equal-regional weighting first satisfies the threshold at the scheduled
	monitor after completed Adam step 1500, followed by the global formulation
	at step 1800 and ARRB at step 2000.
	
	The best physical-global monitor values obtained during Adam are
	
	\begin{align}
		\text{Global hard-constrained:}\quad
		&1.626\times10^{-8},
		\\
		\text{Equal regional:}\quad
		&1.532\times10^{-8},
		\\
		\text{ARRB:}\quad
		&6.997\times10^{-9}.
	\end{align}
	
	Thus, for this seed, the lower final residual achieved by ARRB is already
	apparent at the Adam checkpoint stage and is not produced solely by the
	subsequent common L-BFGS refinement.
	
	Because only one random seed is reported, the numerical differences in
	\cref{tab:ablation} should be interpreted as a matched case study rather
	than estimates of expected performance over random initialization and
	sampling variability.
	
	\begin{figure}[htbp]
		
		\centering
		
		\begin{subfigure}[t]{0.49\textwidth}
			\centering
			\safeincludegraphics[width=\linewidth]
			{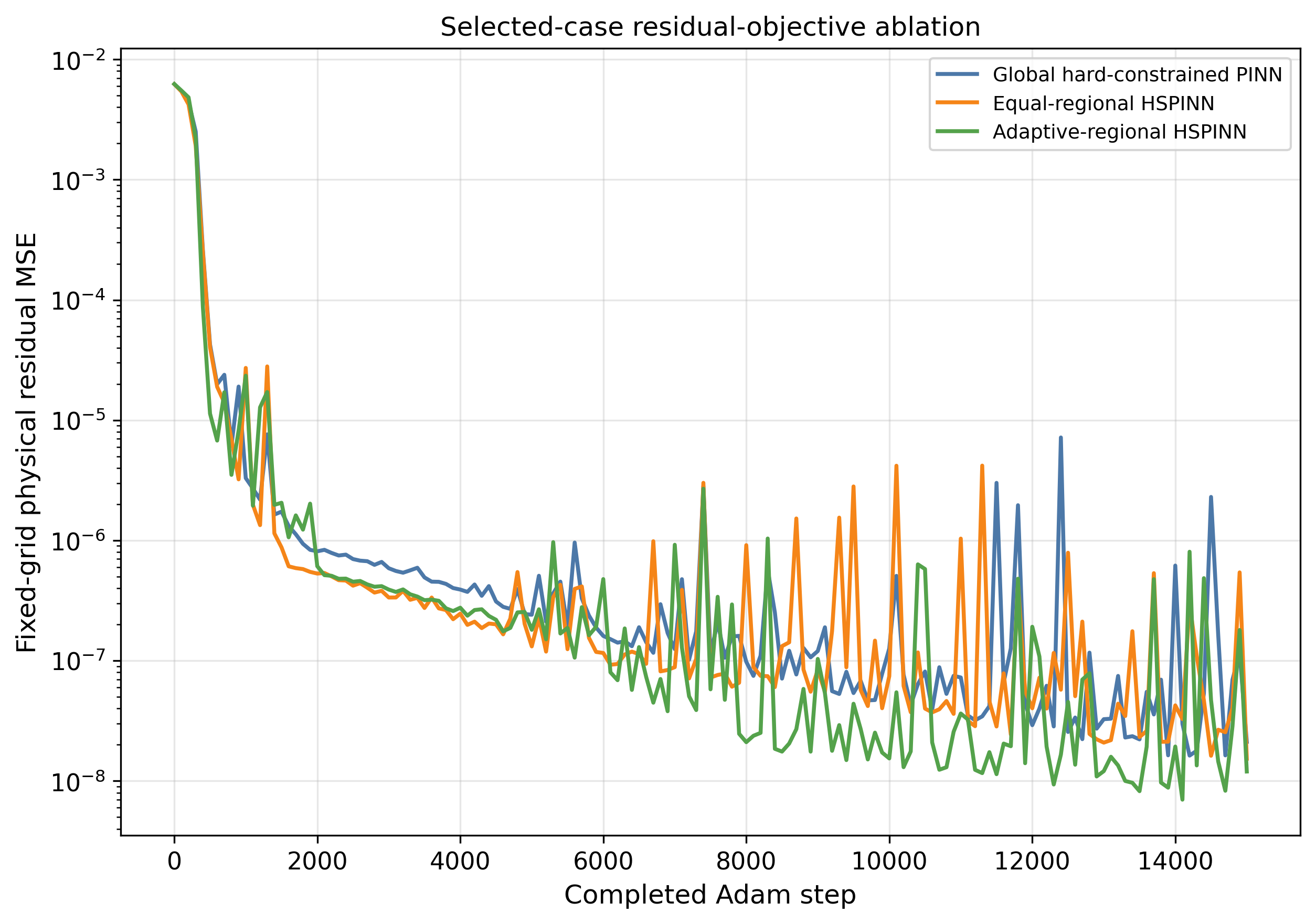}
			{Ablation convergence}
			\caption{Fixed-grid convergence.}
			\label{fig:ablation_convergence}
		\end{subfigure}
		\hfill
		\begin{subfigure}[t]{0.49\textwidth}
			\centering
			\safeincludegraphics[width=\linewidth]
			{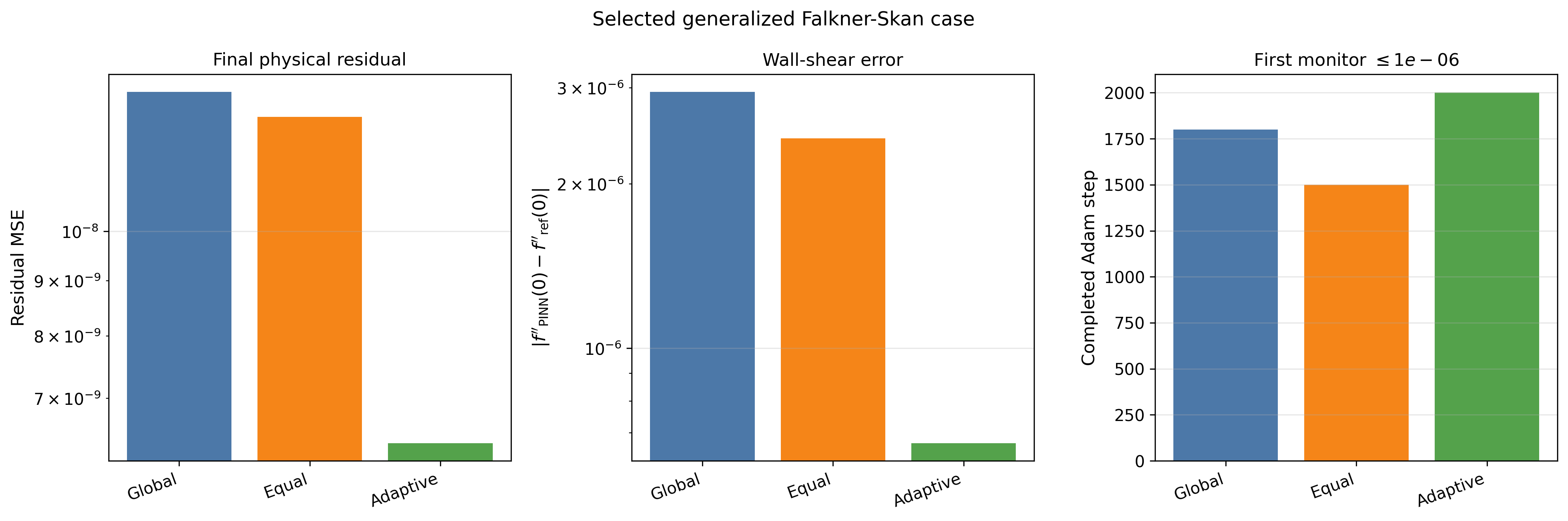}
			{Ablation metrics}
			\caption{Final accuracy metrics.}
			\label{fig:ablation_metrics}
		\end{subfigure}
		
		\caption{
			Matched residual-objective ablation for the selected
			\((\beta_0,\beta_1)=(0.75,0.50)\) Falkner--Skan case.
		}
		
		\label{fig:ablation}
		
	\end{figure}
	
	\Cref{fig:ablation} reinforces the distinction between early threshold
	attainment and final approximation accuracy.
	
	
	\subsection{L-BFGS refinement and rollback}
	\label{subsec:rollback_results}
	For every formulation, L-BFGS minimizes the physical global residual on the stage-specific 2000- or 4000-point grid, whereas acceptance or rollback is determined independently from the 800-point fixed monitor.
	
	For the selected adaptive case, the monitor evolves as
	
	\begin{equation}
		6.997\times10^{-9}
		\rightarrow
		6.801\times10^{-9}
		\rightarrow
		6.798\times10^{-9},
	\end{equation}
	
	so both L-BFGS stages are accepted.
	
	For Blasius,
	
	\begin{equation}
		1.281\times10^{-7}
		\rightarrow
		5.187\times10^{-8}
		\rightarrow
		5.180\times10^{-8},
	\end{equation}
	
	and both stages are again accepted.
	
	For Pohlhausen, the first stage improves the monitor from
	
	\begin{equation}
		8.318\times10^{-8}
	\end{equation}
	
	to
	
	\begin{equation}
		7.234\times10^{-8}.
	\end{equation}
	
	The second stage produces
	
	\begin{equation}
		7.237\times10^{-8},
	\end{equation}
	
	which is slightly larger than the pre-stage value. The second stage is
	therefore rejected and the first-stage model is restored.
	
	In the selected-case ablation, the global baseline improves from
	
	\begin{equation}
		1.626\times10^{-8}
	\end{equation}
	
	to
	
	\begin{equation}
		1.469\times10^{-8}
	\end{equation}
	
	during its first L-BFGS stage, while its second stage is rejected.
	
	The equal-regional formulation similarly improves from
	
	\begin{equation}
		1.532\times10^{-8}
	\end{equation}
	
	to
	
	\begin{equation}
		1.372\times10^{-8}
	\end{equation}
	
	in the first stage; its second stage yields
	
	\begin{equation}
		1.373\times10^{-8}
	\end{equation}
	
	and is rejected.
	
	These outcomes demonstrate the role of the independent rollback criterion:
	a candidate quasi-Newton stage is retained only when it does not degrade
	the prescribed fixed-grid physical-global monitor, even if optimization on
	the stage-specific L-BFGS grid has completed successfully.
	
	\subsection{Observed optimization times}
	\label{subsec:runtime}
	
	The reported timing variable is measured from immediately before the first
	Adam update through completion of the L-BFGS stages and final training-grid
	loss evaluation, but before the subsequent 5000-point evaluation-grid
	post-processing metrics and hard-boundary diagnostics.
	
	The measured CPU optimization wall times for the adaptive selected,
	Blasius, and Pohlhausen cases are approximately
	
	\begin{equation}
		376.04~\mathrm{s},
		\qquad
		475.02~\mathrm{s},
		\qquad
		421.67~\mathrm{s},
	\end{equation}
	
	respectively.
	
	For the selected-case ablation, the global and equal-regional runs require
	approximately
	
	\begin{equation}
		511.70~\mathrm{s}
		\qquad
		\text{and}
		\qquad
		521.01~\mathrm{s},
	\end{equation}
	
	respectively. The adaptive selected run requires \(376.04~\mathrm{s}\).
	
	These CPU wall times are measurements from individual runs on the reported
	publication device. They are provided for reproducibility and workload
	context only. Because the experiments were not repeated specifically for
	timing statistics and because the compared methods can follow different
	optimization trajectories, the single-run values are not interpreted as
	evidence of systematic computational-speed differences.
	\section{Discussion}
	\label{sec:discussion}
	
	The numerical experiments show that the boundary-admissible representation can recover the tested generalized Falkner--Skan solutions with small wall-shear, profile, and governing-equation residual errors under the common computational setting summarized in \cref{tab:configuration}. The displayed zero boundary errors arise from the analytical trial construction rather than from learned boundary fitting. Consequently, interior residual accuracy, differentiated-profile errors, and recovery of the equation-determined wall shear provide the more informative numerical accuracy measures.
	
	The roles of the hard-boundary representation and ARRB should be distinguished. Exact boundary admissibility removes competition between a governing-equation objective and boundary-condition penalty terms. ARRB acts after this reduction and addresses a different issue: unequal parameter-gradient contributions produced by spatial portions of the same governing-equation residual. The results should therefore not be interpreted as evidence that hard constraints themselves are new, nor as an ablation of hard versus soft boundary enforcement.
	
	Across the three benchmark problems, the quantitative results in \cref{tab:case_results} show that the selected favorable-pressure-gradient case gives the smallest physical global residual MSE, \(6.357\times10^{-9}\), and the smallest maximum residual, \(7.540\times10^{-4}\). Its wall-shear error is \(6.691\times10^{-7}\), while its relative shear-profile error is \(4.188\times10^{-5}\). For Blasius, the predicted wall shear is \(0.3320572256\), compared with the finite-domain BVP reference \(0.3320573362\), corresponding to an absolute difference of \(1.106\times10^{-7}\). For Pohlhausen, the predicted wall shear is \(1.1546922871\), which differs by \(8.251\times10^{-6}\) from the analytical semi-infinite-domain value \(2/\sqrt3\). The Pohlhausen case also has the largest maximum pointwise residual among the three benchmark runs, \(3.406\times10^{-3}\).
	
	The profile comparisons in \cref{fig:profile_comparison}, including the velocity and shear subfigures in \cref{fig:velocity_profiles,fig:shear_profiles}, show close agreement between the PINN and the corresponding numerical reference solutions over the computational interval. The reference-profile deviations from Blasius in \cref{fig:velocity_deviation} further illustrate how the selected favorable-pressure-gradient and Pohlhausen cases differ from the zero-pressure-gradient benchmark, providing context for the substantially larger wall-shear values observed in those cases.
	
	The Pohlhausen comparison requires particular care. The analytical solution obeys \(f'(\eta)\rightarrow1\) only as \(\eta\rightarrow\infty\), whereas the PINN and finite-domain BVP impose \(f'(12)=1\) exactly. Consequently, the analytical and finite-domain problems are not mathematically identical at finite \(L\). The analytical wall shear is therefore used as an independent exact benchmark, while the reported discrete profile errors are computed against the BVP posed on the same finite interval as the PINN. The implementation additionally computes the finite-domain BVP wall shear as a truncation diagnostic.
	
	For the selected case, the pointwise diagnostics in \cref{fig:selected_diagnostics}, with profile errors shown in \cref{fig:selected_error} and residual magnitude shown in \cref{fig:selected_residual}, indicate that the small aggregate error measures are not produced by a uniformly exact approximation at every point. These plots therefore complement the scalar residual MSE and relative \(L_2\) errors by revealing the spatial distribution of the remaining approximation error.
	
	A central feature of the method is the separation between physical residual weighting and adaptive optimization weighting. The physical global residual in \cref{eq:global_loss} is reconstructed from regional residual means using region-length fractions rather than stochastic collocation counts. Therefore, the prescribed physical weighting does not change when the integer number of samples assigned to a region changes. The final regional residual MSEs and restored adaptive coefficients in \cref{tab:regional_results} show that the regional residual levels and the adaptive coefficients need not follow the same ordering, consistent with the fact that the coefficients are constructed from proxy-gradient information rather than directly from residual magnitude.
	
	The adaptive coefficients require a separate interpretation. For the selected case, the coefficients restored with the best Adam checkpoint are
	
	\begin{equation}
		(0.7276,\ 0.1816,\ 0.0908).
	\end{equation}
	
	These values arise from safeguarded inverse proxy-gradient norms. A larger \(\lambda_r\) therefore indicates that the associated regional loss would otherwise have a comparatively smaller contribution to the chosen proxy gradient norm; it does not establish a hierarchy of physical importance. Furthermore, because
	\[
	w_r^{\mathrm{eff}}
	=
	\alpha\omega_r+(1-\alpha)\lambda_r,
	\]
	the reported \(\lambda_r\) values are not the complete regional coefficients in the Adam objective. For the selected checkpoint, the effective wall, middle, and tail coefficients are approximately
	
	\begin{equation}
		(0.4888,\ 0.2991,\ 0.2121).
	\end{equation}
	
	The deterministic convergence history and complete adaptive-coefficient trajectory in \cref{fig:monitor} provide additional context for these restored values. In particular, the plotted coefficient history contains all adaptive updates over the Adam run, whereas the coefficients used for the reported restored model are those associated with the best Adam checkpoint. The figure also shows that the lower residual level achieved by the selected ARRB run is already established during Adam rather than being produced solely by the subsequent quasi-Newton refinement.
	
	The matched single-seed ablation in \cref{tab:ablation} provides evidence of improved final accuracy from adaptive weighting in the reported optimization trajectory. Relative to the global hard-constrained formulation, ARRB reduces the final residual MSE by approximately \(52.81\%\) and the wall-shear error by approximately \(77.30\%\). Relative to equal-regional weighting, the corresponding reductions are approximately \(50.24\%\) and \(72.37\%\). ARRB also gives the smallest maximum residual and relative shear-profile error among the three objectives for this seed. These differences are summarized visually in \cref{fig:ablation}, with the convergence histories in \cref{fig:ablation_convergence} and the final accuracy metrics in \cref{fig:ablation_metrics}.
	
	ARRB does not, however, attain the monitored \(10^{-6}\) threshold first. Equal-regional weighting first satisfies the criterion at the scheduled monitor after completed Adam step 1500, followed by the global formulation at step 1800 and ARRB at step 2000, as reported in \cref{tab:ablation} and illustrated in \cref{fig:ablation_convergence}. Because monitoring occurs only at prescribed intervals, these values should not be interpreted as the exact optimizer iterations at which the continuously evolving networks first crossed the threshold. The observed advantage of ARRB in this experiment therefore concerns final approximation accuracy rather than uniformly faster early optimization.
	
	The best Adam physical-global monitor for ARRB is \(6.997\times10^{-9}\), compared with \(1.626\times10^{-8}\) for the global baseline and \(1.532\times10^{-8}\) for equal-regional weighting. Thus, the lower residual obtained in the reported ARRB run is already established during the Adam stage and is not generated solely by L-BFGS refinement.
	
	The deterministic L-BFGS results also illustrate the purpose of the rollback mechanism. Although each L-BFGS stage minimizes the physical global residual on its own 2000- or 4000-point uniform grid, acceptance is based on the independent 800-point monitor specified in \cref{tab:configuration}. The second Pohlhausen stage, together with the second stages of the global and equal-regional ablation baselines, is rejected because the corresponding monitor deteriorates slightly. This separation reduces the risk of accepting a refinement that improves the optimization grid while degrading the prescribed monitoring grid.
	
	Several limitations remain. First, the residual-objective ablation uses a single random seed. Multi-seed matched experiments are therefore required to quantify sensitivity to neural initialization and stochastic collocation sampling and to determine whether the observed accuracy differences persist statistically.
	
	Second, the regional interfaces at \(\eta=3\) and \(\eta=8\), summarized together with the other computational choices in \cref{tab:configuration}, are prescribed rather than learned or optimized. The current implementation supports alternative interfaces and recomputes both physical region-length fractions and largest-remainder sample counts accordingly, but a systematic partition-sensitivity study has not yet been reported.
	
	Third, the adequacy of \(L=12\) should be re-examined for slower-decaying, adverse-pressure-gradient, and near-separation regimes. This is particularly relevant because the far-field condition is represented by an exact finite-domain endpoint condition.
	
	Fourth, the current ablation compares three residual objectives while retaining the same hard-boundary representation. It therefore isolates the residual-weighting mechanism but does not constitute a direct comparison with a conventional soft-boundary PINN.
	
	Finally, the SciPy BVP tolerance controls the numerical solver's convergence criterion and should not be interpreted as an independent rigorous error certificate for every reported reference quantity. Mesh, solver-tolerance, and domain-length sensitivity should be examined when substantially higher reference precision is required.
	
	
	\section{Conclusions}
	\label{sec:conclusions}
	
	This study combines a Falkner--Skan-specific hard-constrained neural
	representation with adaptive regional residual balancing. The analytical
	trial form satisfies
	
	\begin{equation}
		f(0)=0,
		\qquad
		f'(0)=0,
		\qquad
		f'(L)=1
	\end{equation}
	
	identically, apart from floating-point roundoff, while preserving the
	freedom required for \(f''(0)\) and \(f(L)\) to be determined by the
	governing equation.
	
	ARRB partitions the governing-equation residual into wall, middle, and
	tail regions and balances their proxy-gradient contributions during Adam
	training using safeguarded EMA-smoothed inverse-gradient coefficients. A
	separate region-length-weighted physical residual is retained for physical
	global evaluation, deterministic checkpoint monitoring, L-BFGS
	refinement, rollback decisions, and final residual assessment. The
	adaptive coefficients therefore modify optimization conditioning without
	replacing the underlying physical region-length weighting.
	
	For the selected favorable-pressure-gradient case
	\((\beta_0,\beta_1)=(0.75,0.50)\), the method predicts
	
	\begin{equation}
		f''(0)=0.8997161394,
	\end{equation}
	
	compared with the finite-domain numerical BVP reference
	
	\begin{equation}
		f''(0)=0.8997168085.
	\end{equation}
	
	The resulting wall-shear error is
	\(6.691\times10^{-7}\), the 5000-point physical residual MSE is
	\(6.357\times10^{-9}\), the maximum absolute residual is
	\(7.540\times10^{-4}\), and the relative shear-profile error is
	\(4.188\times10^{-5}\).
	
	In the reported matched single-seed residual-objective ablation, ARRB
	gives the smallest final residual MSE, maximum pointwise residual,
	wall-shear error, and relative shear-profile error among the three tested
	objectives. Relative to the global hard-constrained formulation, the
	residual MSE and wall-shear error are reduced by approximately
	\(52.81\%\) and \(77.30\%\), respectively. Relative to equal-regional
	weighting, the corresponding reductions are approximately \(50.24\%\)
	and \(72.37\%\).
	
	These improvements should be interpreted as results for the reported
	matched optimization trajectory rather than as statistical performance
	estimates. ARRB also does not reach the prescribed monitored
	early-convergence threshold first: equal-regional weighting reaches the
	criterion at an earlier scheduled monitor evaluation. The present results
	therefore support ARRB primarily as an accuracy-oriented
	residual-conditioning mechanism rather than as a guarantee of faster
	optimization.
	
	Future work should quantify multi-seed variability, investigate
	sensitivity to the prescribed regional interfaces and domain truncation,
	extend the tests to adverse-pressure-gradient and near-separation regimes,
	compare directly against conventional soft-boundary PINNs, and explore
	adaptive or automatically determined regional decompositions and multi-domain extensions.
	
	
	%
	%
	

\end{document}